\documentclass[aps,pre,notitlepage,reprint,onecolumn,superscriptaddress]{revtex4-1}

\usepackage{epsfig}
\usepackage{natbib}

\usepackage{xspace}
\usepackage{xcolor}
\usepackage{graphicx}
\usepackage{amssymb}
\usepackage{cancel}

\usepackage{lineno}
\usepackage{amsmath}
\usepackage{stackrel}
\usepackage{appendix}
\usepackage{url}
 \usepackage{enumitem}
 \usepackage{hyperref}
\usepackage{xcolor}
\usepackage{array}
\usepackage{booktabs}
\usepackage{multirow}

\hypersetup{
	colorlinks=true,
	linkcolor=blue,
	citecolor=blue,
	urlcolor=blue
}
\RequirePackage[normalem]{ulem} 
\RequirePackage{color}\definecolor{RED}{rgb}{1,0,0}\definecolor{BLACK}{rgb}{0,0,1} 
\providecommand{\DIFaddtex}[1]{{\protect\color{black}{#1}}} 
\providecommand{\DIFdeltex}[1]{{\protect\color{red}\sout{#1}}}                      
\providecommand{\DIFaddbegin}{} 
\providecommand{\DIFaddend}{} 
\providecommand{\DIFdelbegin}{} 
\providecommand{\DIFdelend}{} 
\providecommand{\DIFaddFL}[1]{\DIFadd{#1}} 
\providecommand{\DIFdelFL}[1]{\DIFdel{#1}} 
\providecommand{\DIFaddbeginFL}{} 
\providecommand{\DIFaddendFL}{} 
\providecommand{\DIFdelbeginFL}{} 
\providecommand{\DIFdelendFL}{} 
\providecommand{\DIFadd}[1]{\texorpdfstring{\DIFaddtex{#1}}{#1}} 
\providecommand{\DIFdel}[1]{\texorpdfstring{\DIFdeltex{#1}}{}} 
\newcommand{\DIFscaledelfig}{0.5}
\RequirePackage{settobox} 
\RequirePackage{letltxmacro} 
\newsavebox{\DIFdelgraphicsbox} 
\newlength{\DIFdelgraphicswidth} 
\newlength{\DIFdelgraphicsheight} 
\LetLtxMacro{\DIFOincludegraphics}{\includegraphics} 
\newcommand{\DIFaddincludegraphics}[2][]{{\color{black}\fbox{\DIFOincludegraphics[#1]{#2}}}} 
\newcommand{\DIFdelincludegraphics}[2][]{
\sbox{\DIFdelgraphicsbox}{\DIFOincludegraphics[#1]{#2}}
\settoboxwidth{\DIFdelgraphicswidth}{\DIFdelgraphicsbox} 
\settoboxtotalheight{\DIFdelgraphicsheight}{\DIFdelgraphicsbox} 
\scalebox{\DIFscaledelfig}{
\parbox[b]{\DIFdelgraphicswidth}{\usebox{\DIFdelgraphicsbox}\\[-\baselineskip] \rule{\DIFdelgraphicswidth}{0em}}\llap{\resizebox{\DIFdelgraphicswidth}{\DIFdelgraphicsheight}{
\setlength{\unitlength}{\DIFdelgraphicswidth}
\begin{picture}(1,1)
\thicklines\linethickness{2pt} 
{\color[rgb]{1,0,0}\put(0,0){\framebox(1,1){}}}
{\color[rgb]{1,0,0}\put(0,0){\line( 1,1){1}}}
{\color[rgb]{1,0,0}\put(0,1){\line(1,-1){1}}}
\end{picture}
}\hspace*{3pt}}} 
} 
\LetLtxMacro{\DIFOaddbegin}{\DIFaddbegin} 
\LetLtxMacro{\DIFOaddend}{\DIFaddend} 
\LetLtxMacro{\DIFOdelbegin}{\DIFdelbegin} 
\LetLtxMacro{\DIFOdelend}{\DIFdelend} 
\DeclareRobustCommand{\DIFaddbegin}{\DIFOaddbegin \let\includegraphics\DIFaddincludegraphics} 
\DeclareRobustCommand{\DIFaddend}{\DIFOaddend \let\includegraphics\DIFOincludegraphics} 
\DeclareRobustCommand{\DIFdelbegin}{\DIFOdelbegin \let\includegraphics\DIFdelincludegraphics} 
\DeclareRobustCommand{\DIFdelend}{\DIFOaddend \let\includegraphics\DIFOincludegraphics} 
\LetLtxMacro{\DIFOaddbeginFL}{\DIFaddbeginFL} 
\LetLtxMacro{\DIFOaddendFL}{\DIFaddendFL} 
\LetLtxMacro{\DIFOdelbeginFL}{\DIFdelbeginFL} 
\LetLtxMacro{\DIFOdelendFL}{\DIFdelendFL} 
\DeclareRobustCommand{\DIFaddbeginFL}{\DIFOaddbeginFL \let\includegraphics\DIFaddincludegraphics} 
\DeclareRobustCommand{\DIFaddendFL}{\DIFOaddendFL \let\includegraphics\DIFOincludegraphics} 
\DeclareRobustCommand{\DIFdelbeginFL}{\DIFOdelbeginFL \let\includegraphics\DIFdelincludegraphics} 
\DeclareRobustCommand{\DIFdelendFL}{\DIFOaddendFL \let\includegraphics\DIFOincludegraphics} 
\RequirePackage{listings} 
\RequirePackage{color} 
\lstdefinelanguage{DIFcode}{ 
  moredelim=[il][\color{red}\sout]{\%DIF\ <\ }, 
  moredelim=[il][\color{black}\uwave]{\%DIF\ >\ } 
} 
\lstdefinestyle{DIFverbatimstyle}{ 
	language=DIFcode, 
	basicstyle=\ttfamily, 
	columns=fullflexible, 
	keepspaces=true 
} 
\lstnewenvironment{DIFverbatim}{\lstset{style=DIFverbatimstyle}}{} 
\lstnewenvironment{DIFverbatim*}{\lstset{style=DIFverbatimstyle,showspaces=true}}{} 

\begin{document}
\title{The Thermodynamic Limit of Extreme First-Passage Times}

\author{Talia Baravi}
\affiliation{Department of Physics, Institute of Nanotechnology and Advanced Materials, Bar-Ilan University, Ramat Gan 52900, Israel}

\author{Eli Barkai}
\affiliation{Department of Physics, Institute of Nanotechnology and Advanced Materials, Bar-Ilan University, Ramat Gan 52900, Israel}  

\begin{abstract}

The statistics of the slowest first-passage time among a large population of $N$ searchers is crucial for determining the completion time of many stochastic processes. Classical extreme-value theory predicts that for diffusing particles in a finite domain of size $L$, the slowest first passage time follows a Gumbel distribution, but a Fréchet distribution in an infinite domain. Here, we study the physically relevant thermodynamic limit where both $N$ and $L$ diverge while the density $\rho = N/L$ remains constant.
We obtain an explicit solution for the extreme value in the thermodynamic limit, which recovers the Fréchet and Gumbel distributions in the low- and high-density limits, respectively, and reveals new, nontrivial behavior at intermediate densities. We then extend the framework to compact diffusion on fractal domains, \DIFaddbegin \DIFadd{such as the Sierpiński gasket, }\DIFaddend showing that the walk dimension $d_w$ and fractal dimension $d_f$ control the extreme-value statistics via geometry-dependent scaling. The theory yields the full set of moments and finite-density corrections, providing a unified description of slowest-arrival times in confined Euclidean and fractal media.

\end{abstract}

\maketitle

\section{Introduction}
Extreme-value (EV) theory seeks the limiting laws that describe the maxima (or minima) out of a set of $N$ random variables. It is a foundational statistical framework with wide-ranging applications in fields like climatology, finance, and engineering for predicting rare events \DIFaddbegin \DIFadd{\mbox{
\cite{schuss2019redundancy,majumdar2024statistics,lawley2023slowest,guerin2016mean,jenkinson1955frequency,novak2011extreme,majumdar2002extreme,fortin2015applications,biroli2025stronglycorrelatedstochasticsystems,matsinos2024extreme,holl2020extreme,albeverio2006extreme,embrechts2013modelling,huang2025first,bouchaud1997universality,flandoli2025extreme}}\hskip0pt
}\DIFaddend . In classical EV theory, the asymptotic behavior of the maximum out of $N$ independent and identically distributed (i.i.d.) random variables depends critically on the tail of the parent distribution. More precisely, EV theory distinguishes three universal limiting distributions: the Gumbel, Fréchet, and Weibull families, corresponding respectively to light-tailed, heavy-tailed, and bounded-support distributions \cite{gumbel1958statistics,fisher1928limiting,gnedenko1943distribution,majumdar2020extreme}. In practice, however, direct application of these universal laws is often complicated by finite-sample effects: convergence to the asymptotic limits can be slow, and finite-size corrections that dominate empirically relevant ranges can be substantial \cite{hall1979rate,gyorgyi2008finite,majumdar2024decorrelation,zarfaty2022discrete,zarfaty2021accurately,mikosch2020gumbel, oshanin2013anomalous}. In applied work, a common approach is to use the generalized EV distribution \cite{fn:GEVcomment}, a flexible three-parameter framework that combines the Gumbel, Fréchet, and Weibull distributions \cite{jenkinson1955frequency, de2018superstatistical,kotz2000extreme,singh1998generalized,raynal2021general,tsiftsi2018extreme}. While useful for fitting empirical data, the generalized EV approach is often applied ad-hoc, without a direct link to the underlying physical process generating the data.

The classical framework is built on a single large parameter: the number $N$ of i.i.d. random variables, drawn from a common parent distribution. A different kind of complexity arises in physical systems where the parent distribution is parameterized by a macroscopic physical scale, such as a system of size $L$. When $L$ grows in proportion to the sample size $N$, the problem moves beyond the scope of classical EV theory. This scenario gives rise to a thermodynamic limit, where both $N\rightarrow \infty$ and $L\rightarrow \infty$ while their ratio, the density $\rho=N/L$, remains constant. In this limit, the statistics of extremes can deviate significantly from the classical EV laws, potentially undergoing a crossover from one class to another. To explore this general problem, we investigate a canonical example: the extreme statistics of first-passage times (FPTs) for a population of diffusing particles.

\par In stochastic systems involving diffusive motion, the time it takes for a particle to reach a target for the first time, known as the FPT, is a central quantity with broad relevance across physics, chemistry, and biology  \DIFaddbegin \DIFadd{\mbox{
\cite{redner2001guide,TargetSearch2024,benichou2014first,godec2016universal,bray2013persistence,meyer2011universality,guerin2016mean,grebenkov2023boundary,scher2023escape,singh2025sokoban}}\hskip0pt
}\DIFaddend . In the study of multiple independent i.e., non-interacting searchers, interest often shifts from the FPT of a single particle to the extreme FPTs: the fastest or the slowest events among a population. The statistics of the fastest FPT have been extensively studied and shown to play key roles in rapid detection or triggering phenomena \DIFaddbegin \DIFadd{\mbox{
\cite{lawley2024competition,grebenkov2022first,basnayake2019asymptotic,lawley2020distribution,lawley2020extreme,linn2022extreme,lawley2020extreme,ellettari2025rare,meerson2015mortality,franke2012survival}}\hskip0pt
. These studies include different formulations of diffusion-limited search processes, such as a swarm of independent particles searching for a fixed target or a single walker diffusing among stationary traps. }\DIFaddend The slowest FPT, in contrast, governs the overall completion time in a wide range of systems  \DIFaddbegin \DIFadd{\mbox{
\cite{schuss2019redundancy,mejia2011first,madrid2020competition,basnayake2019fastest,bao2006last}}\hskip0pt
}\DIFaddend : for instance, menopause onset has been modeled as the time at which the last of a finite ovarian follicle reserve becomes inactive~\cite{lawley2023slowest}; in intracellular processes, degradation or signal transduction may require all particles to be cleared or absorbed~\cite{grebenkov2022reversible}; and in distributed computing or molecular assembly, a process may not complete until every component has reached its destination or reacted~\cite{ghodsi1991performance,dey2019dissipative}. In this paper we will focus on the slowest events, though our methods can be applied also to the opposite situation.

    Let us consider $N$ i.i.d. random variables $t_1,t_2,...,t_N$ representing the FPTs of $N$ Brownian particles undergoing overdamped Langevin dynamics, such as Brownian motion in free space or within a confined domain of size $L$. The slowest FPT is then given by
$
    \mathcal{T} :=\text{max}\{t_1,t_2,...,t_N\}\, .
$
Understanding the distribution of $\mathcal{T}$ is crucial in settings where a process completes only once all particles have reached a target.
In the context of diffusive systems, if the domain is infinite or unbounded, the single-particle FPT typically exhibits a power-law tail \cite{schrodinger1915theorie,majumdar1999persistence,havlin2002diffusion}, therefore the slowest FPT falls in the Fr\'echet class. Conversely, in finite domains, the survival probability decays exponentially for large times, and the slowest FPT converges to a Gumbel distribution as $N\rightarrow \infty$ \cite{fisher1928limiting,gnedenko1943distribution,gumbel1958statistics}. This was demonstrated by Lawley and Johnson \cite{lawley2023slowest}. They showed that the slowest FPT of diffusing particles in bounded domains exhibits logarithmic scaling with \( N \), consistent with Gumbel statistics. A similar Gumbel limit was also observed in systems with constant drift, where the drift induces an exponential cutoff akin to confinement \cite{comtet2020last}.

However, between these two classical regimes lies an intermediate but physically relevant limit: the thermodynamic limit where both the number of particles $N$ and the domain size $L$ grow, while keeping the density $\rho=N/L$ fixed. In this setting, neither the Gumbel nor the Fr\'echet laws alone fully describe the behavior of the slowest FPT. This limit reflects many real-world scenarios where the number of particles scales with system size, such as diffusive transport in crowded biological environments or throughput analysis in parallel search algorithms \cite{ghodsi1991performance,meinecke2017multiscale}.

In this paper, we address this gap by developing a theory for the maximal FPT of $N$ non-interacting Brownian particles in a 1D interval in the finite-density limit. Starting from the exact single-particle survival probability, we derive a universal scaling function for the cumulative distribution of the slowest FPT that captures both the exponential and algebraic decay limits. This approach allows us to describe the full crossover between Gumbel and Fr\'echet statistics as the density is varied, and to compute explicit expressions for the mean and general moments of the slowest FPT.
\par The example of a particle confined in a box serves as a simple illustration of a more general framework for analyzing the extremes of FPTs, provided the underlying search process is compact. Compactness here refers to the property that a random walker returns to its origin with probability one. Building on our recent results for first-passage times of a single particle \cite{baravi2025first}, we now extend the theory to extreme-value statistics in the thermodynamic limit, which will be defined precisely below. Beyond the simple box geometry, we consider diffusion on fractal domains as a prototypical compact process. In this case, the interplay between the walk dimension $d_w$ and the fractal dimension $d_f$ determines the scaling of the survival probability and thus governs the extreme statistics of first-passage times.

\par The paper is organized as follows. In Section~\ref{sec:second}, we review the emergence of Gumbel and Fréchet statistics from the distribution of the slowest FPT in their respective large-\( N \) and large-\( L \) limits. In Section~\ref{sec:thermlim}, we develop the thermodynamic limit by deriving the single-particle survival probability and constructing the scaling form for the maximum FPT at fixed density. Section~\ref{sec:extmom} analyzes the moments of the slowest FPT and compares our predictions with exact numerical results.
Finally, Section~\ref{sec:exten} demonstrates the universality of the framework by applying it to diffusion on a fractal geometry and showing how the same scaling structure governs the model-specific extreme-value crossover.

\section{Fr\'echet and Gumbel's statistics}\label{sec:second}
This section provides a recap of Fr\'echet and Gumbel's statistics in the context of extreme value analysis for first-passage times. Consider $N$ Brownian particles confined within a one-dimensional box, where each particle is initially
positioned at $x_0 > 0$ at time $t = 0$. The particles diffuse independently, and their dynamics are governed by the
diffusion equation
\begin{equation}\label{eq:FP}
\frac{\partial P(x, t)}{\partial t} = D \frac{\partial^2 P(x, t)}{\partial x^2} \, ,
\end{equation}
where $P(x,t)$ is the probability density of a particle being at position $x$ at time $t$, and $D$ is the diffusion constant.
We consider the first-passage time for a particle to exit a finite interval, where escape can occur through either the left or right boundary. The boundaries of the box at $x = 0$ and $x = L$ are absorbing, therefore $P(x = 0, t) = P(x = L, t) = 0$. The solution to this equation is well-known and will be given below. Each particle's escape is governed by a stochastic first-passage time (FPT). The FPTs \( \{t_{i}\} \) are independent and identically distributed (i.i.d.) random variables, where $i=1,2,...,N$. The probability that the maximum FPT \( \mathcal{T} = \max\{t_i\} \) is smaller than a given threshold \(\mathcal{T} \) is denoted by \( Q_N(\mathcal{T}) \). Since the \(\{t_i\}\) are i.i.d. random variables, this probability can be expressed through factorization as \cite{feller1971introduction}
\begin{equation}\label{eq:nparticles1}
    Q_N(\mathcal{T})=\mbox{Prob}(t_1,t_2,...,t_N<\mathcal{T}) = \left[1-S(\mathcal{T})\right]^N  \, ,
\end{equation}
where
\begin{equation}\label{eq:defsurv}
   S(\mathcal{T})=\int_{\mathcal{T}}^{\infty} dt' p(t')  \, ,
\end{equation}
and therefore the PDF of the largest FPT is given by
\begin{equation}\label{eq:defq}
    q_N(\mathcal{T})=\frac{\partial}{\partial \mathcal{T} } Q_N(\mathcal{T})=  N\,p(\mathcal{T})\left[1-S(\mathcal{T})\right]^{N-1}  \, .
\end{equation}
Here, $S(\cdot)$ is the survival probability and \( p(\cdot) \) is the parent PDF of the FPT for a single particle, which is obtained from Eq.\,(\ref{eq:FP}), see below. The survival probability $S(t)$ is the probability that a particle diffusing in the interval $[0, L]$ remains inside the domain without hitting the absorbing boundaries up to time $t$. While the expression for \( Q_N(\mathcal{T}) \) provides the exact cumulative distribution of the slowest escape time \( \mathcal{T} \), it requires knowledge of \( p(t) \) for all \( t \geq 0 \). 

Moreover, since \( \int_\mathcal{T}^\infty p(t) \, dt < 1 \), the term inside the bracket in Eq.\,(\ref{eq:defq}), \( 1 - \int_\mathcal{T}^\infty p(t) \, dt \), is strictly less than one. Consequently, in the limit \( N \to \infty \), the expression for \( Q_N(\mathcal{T}) \) is supposedly expected to converge to zero for any finite \( \mathcal{T} \). To obtain a finite limit, $\mathcal{T} $ is rescaled to align with the asymptotic behavior of the parent distribution.

In the limit of large $N$, the statistics of the slowest FPT converge to a limiting distribution, as established in classic results of extreme value theory \cite{fisher1928limiting,gnedenko1943distribution,majumdar2020extreme,gnedenko1943distribution,leadbetter2012extremes}. This is because when the number of particles $N$ is large, the slowest first-passage time $\mathcal{T}$ is also large. To obtain a finite limiting distribution, a rescaling of the time variable is necessary
\begin{equation}\label{eq:limitf}
\lim_{\mathcal{T}, N \to \infty} Q_N(\mathcal{T}) \to F(z),
\end{equation}
where \( F(z) \) is the limiting cumulative distribution. The scaled variable $z$ is fixed and given by 
\begin{equation}\label{eq:defz1}
z = \frac{\mathcal{T} - a_N}{b_N}.
\end{equation}
The specific forms of \( a_N \) and \( b_N \) depend in general on the parent distribution and will be identified later. As we already mentioned, the limiting law depends on the parameters $N$ and $L$, and their ratio.

\begin{figure}[h]
    \centering
    \begin{minipage}[t]{0.49\linewidth}
        \centering
        \includegraphics[width=\linewidth]{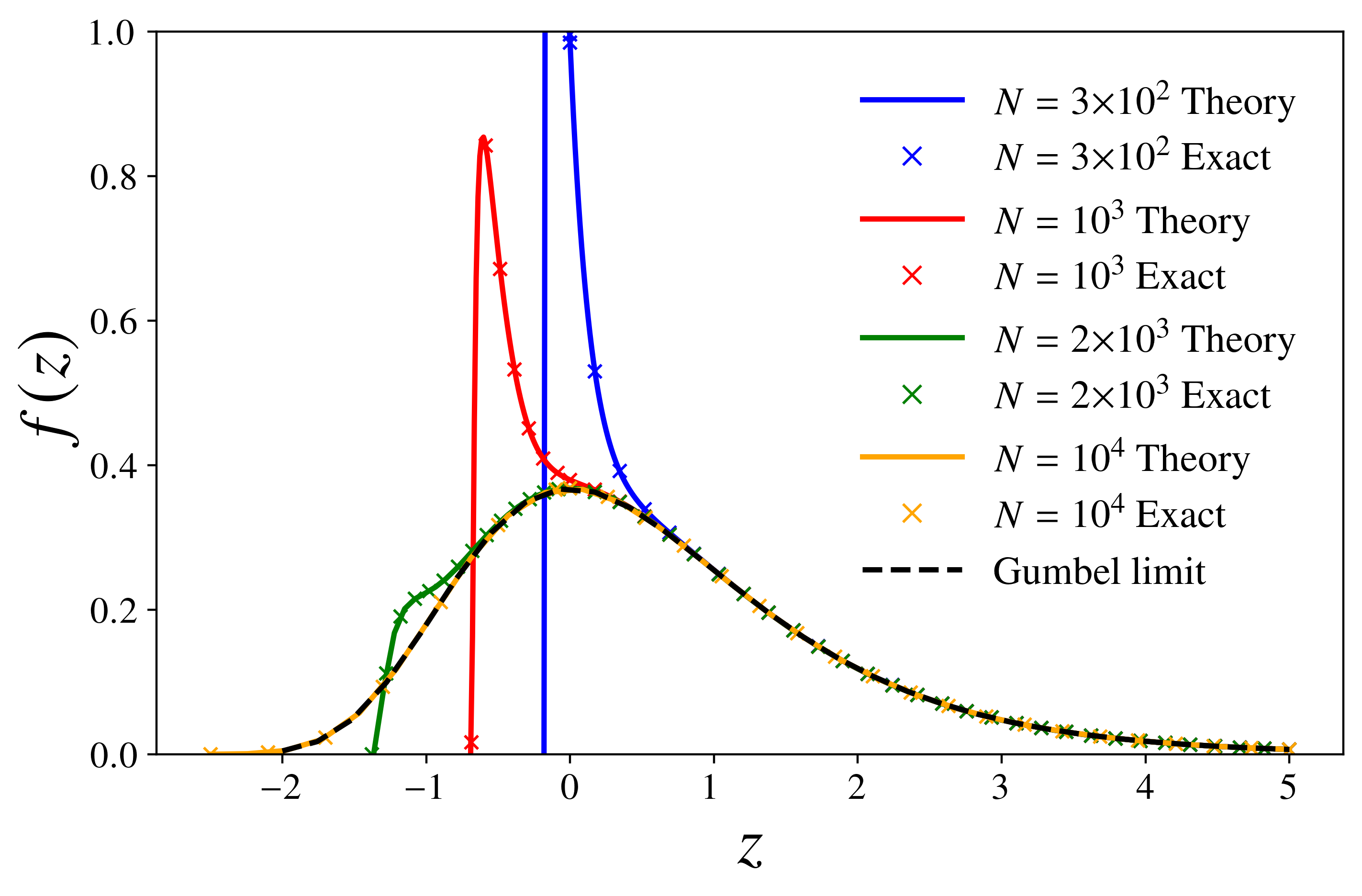}
        \caption{Convergence of the exact PDF of the slowest first-passage time for a Brownian particle following the diffusion Eq.\,(\ref{eq:FP}) to the Gumbel probability density function, given by Eq.\,(\ref{eq:gumbelpdf}), as \( N \) increases. The results are shown for a fixed system size \( L = 1000 \) as a function of $z = 4\,\mathcal{T}/\pi^2L^2  -  \log 2x_0\rho $, see also Appendix \ref{append1a}. The dashed line represents the Gumbel density function, while the symbols denote the exact results and the solid lines the theory developed in this paper, which will be discussed in detail later, see Eqs.\,(\ref{eq:newzdef},\ref{eq:theorys}). Here we used $D=1/2$ and $x_0=1$. As we increase $N$, the convergence to the asymptotic limit is clear, though deviations are visible for small $z$.}\label{fig:comparegumbel}
    \end{minipage}
    \hfill
    \begin{minipage}[t]{0.49\linewidth}
        \centering
        \includegraphics[width=\linewidth]{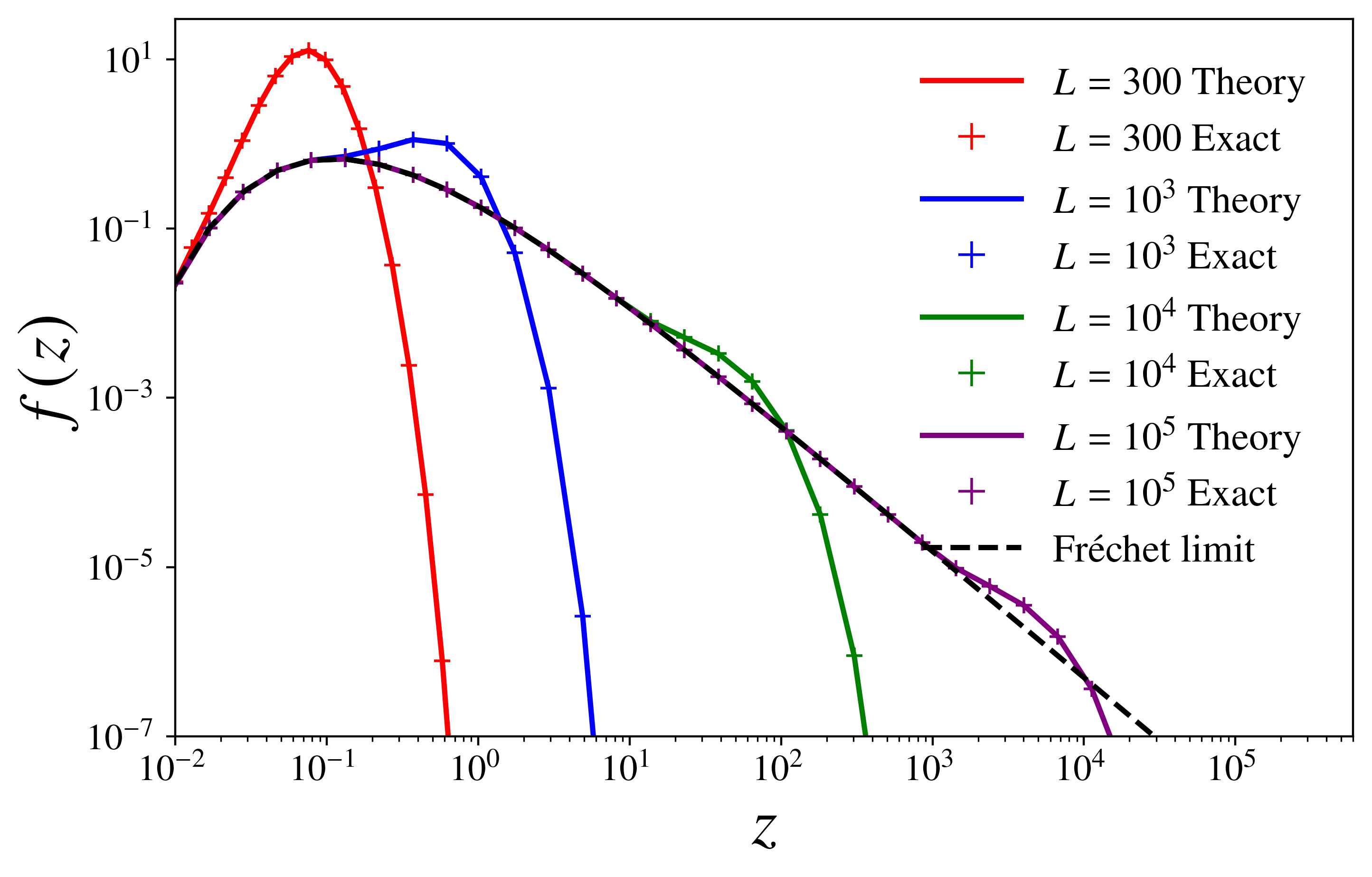}
        \caption{Convergence of the exact PDF of the slowest first-passage time to the Fréchet density function, given by Eq.\,(\ref{eq:frechetpdf}), as the system size \( L \) increases for fixed $N=10^3$. The dashed line represent the Fréchet probability density function, the symbols show the exact numerical solution, and the solid lines denotes the results from our theoretical framework, which will be discussed in detail later in Eqs.\,(\ref{eq:newzdef},\ref{eq:theorys}). The rescaled variable $z = D\,\pi\,\mathcal{T}/x_0^2 N^2$ (see also appendix \ref{append1b}), using parameters $D=1/2$ and $x_0=1$. Notably, the Fr\'echet distribution provides a good fit to the data when $L=10^5$, except at very large values of $z$, while for smaller system sizes, it does not accurately capture the behavior. }\label{fig:comparefresh}
    \end{minipage}
\end{figure}

Two simple laws emerge for the limiting distribution of the EV. Consider a box of finite size $L$, with absorbing boundary at $x=0$. The FPT distribution of a single particle exhibits an exponential tail for large times \cite{redner2001guide,godec2016universal,benichou2014first}. As a result, the slowest escape time among $N$ i.i.d. random variables with exponential tails, converges to the Gumbel distribution as $N\rightarrow \infty$ \cite{gumbel1958statistics,gnedenko1943distribution,eliazar2019poisson}. This Gumbel distribution is a specific limiting form of $F(z)$ in Eq.\,(\ref{eq:limitf}) and is given by
\begin{equation}\label{eq:gumbcdf}
F_{\text{Gumbel}}(z) = \exp\left[- e^{-z}\right] \, ,
\end{equation}
therefore the PDF is
\begin{equation}\label{eq:gumbelpdf}
f_{\text{Gumbel}}(z) = \exp\left[-z - e^{-z}\right].
\end{equation}
For the model under study, with \( L = 1000 \), \( D = 1/2 \), and \( x_0 = 1 \), Fig.\,\ref{fig:comparegumbel} demonstrates the convergence of the exact PDF in Eq.\,(\ref{eq:defq}) to the limiting Gumbel density function as \( N \) increases. The details of the numerical implementation, including the specific values of the rescaling parameters $a_N,b_N$, as well as the procedure for generating the figures, are provided in Appendix \ref{appendix:Gumbel}.

A different limit arises when the system size \( L \to \infty \), while the number of particles \( N \) remains large but finite. In this case, the FPT PDF of a single particle exhibits a well-known power-law tail for large \( t \), given by \cite{redner2001guide}

\begin{equation}\label{eq:freshtail}
p(t) \propto t^{-3/2}, \quad t \to \infty \,.
\end{equation}
This heavy-tailed behavior impacts the statistics of the slowest escape time, for any finite time \( t \). Consequently, in this regime, the maximum \( t \) converges to the Fr\'echet distribution, which is characteristic of random variables with heavy-tailed distributions, such as power laws. The cumulative distribution function of the Fréchet distribution is
\begin{equation}\label{eq:freshcdf}
F_{\text{Fréchet}}(z) = 
\begin{cases} 
0, & z \leq 0, \\
\exp[-z^{-\alpha}], & z > 0,
\end{cases}
\end{equation}
and the PDF is therefore 
\begin{equation}\label{eq:frechetpdf}
f_{\text{Fréchet}}(z) = \alpha z^{-(1+\alpha)} \exp[-z^{-\alpha}], \quad z > 0,
\end{equation}
where \( \alpha > 0 \) is the tail index. For the PDF in Eq.\,(\ref{eq:freshtail}) the corresponding tail index is $\alpha=1/2$. In Fig.\,\ref{fig:comparefresh}, we illustrate the convergence of the exact PDF of the rescaled largest FPT to the Fréchet density function, given by Eq.\,(\ref{eq:frechetpdf}), as the system size \( L \) increases relative to the number of particles \( N \). See Appendix \ref{appendix:Gumbel} for details.

We see that when the system size \( L \) is fixed and \( N \rightarrow \infty\), the statistics are governed by Gumbel's law. In contrast, if \( N \) is large but finite and the system size increases as \( L\rightarrow \infty \), the statistics transition to Fréchet behavior. Since these limits do not commute, they fail to capture the full range of possible extreme-value behavior.

\begin{figure}[h]
  \centering
    \includegraphics[width=0.8\linewidth]{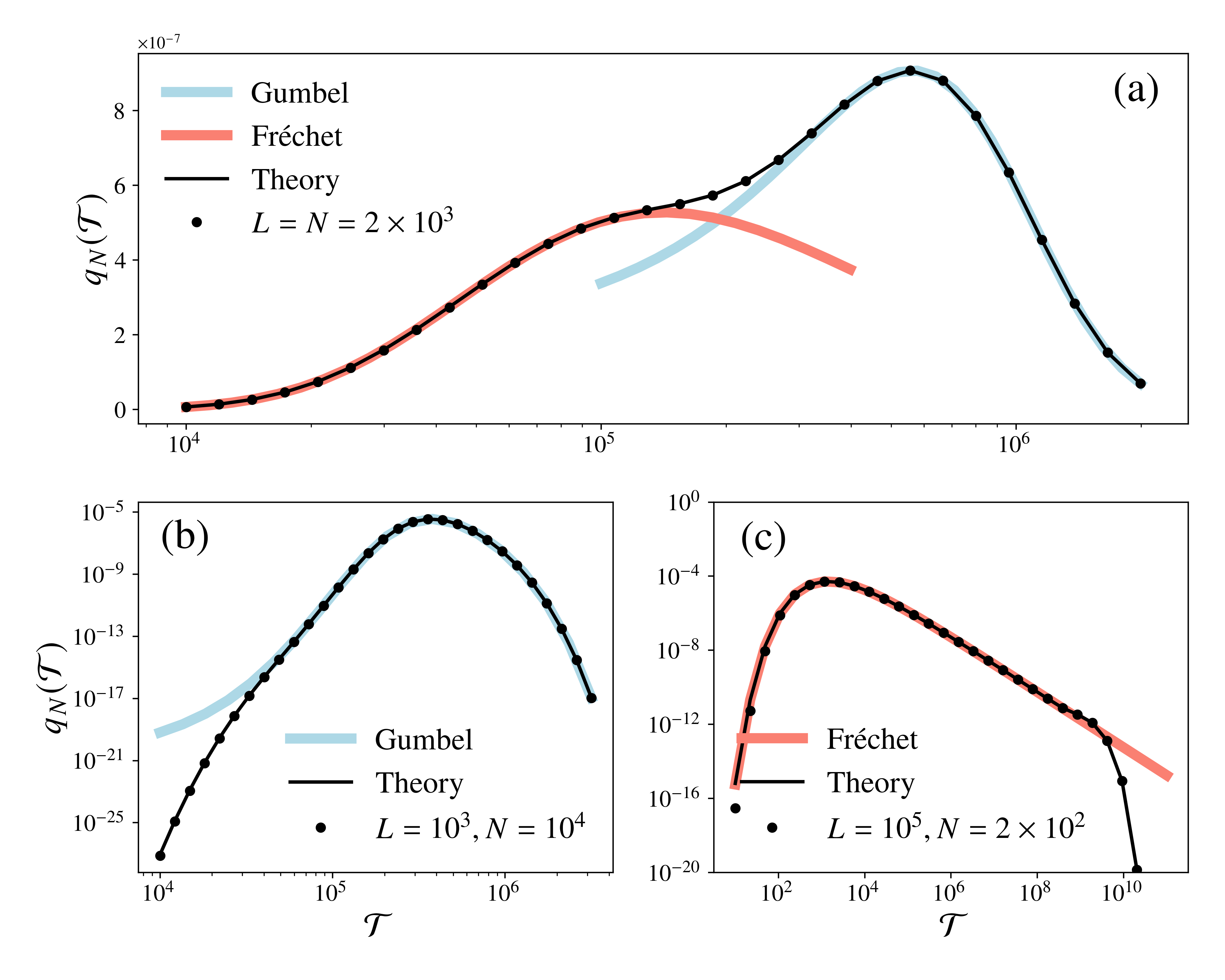}
  \caption{PDF of the slowest FPT $\mathcal{T}$  for different regimes of \( N \) and \( L \) with $x_0=1$. The solid black line represents the theoretical prediction from Eqs.\,(\ref{eq:surv1},\ref{eq:qapprox}), and the symbols denote the exact PDF:  
(a) \( N = L/x_0 \): Comparison with both Gumbel and Fréchet shows that neither limit fully describes the behavior, with Gumbel fitting better at long times and Fréchet at short times.  
(b) High density limit \( N \gg L/x_0 \): The exact results are compared to Gumbel, which provides a good fit for large \( N \) and finite \( L \).  
(c) Low density limit \( N \ll L/x_0 \): The exact results are compared to Fréchet, which accurately captures the behavior in the large system limit with small \( N \), provided that $\mathcal{T}$ is not too large. In all figures the thermodynamic limit we studied in Eqs.\,(\ref{eq:surv1},\ref{eq:qapprox}) matches nicely the exact results, with no fitting. }\label{fig:comparevst}
\end{figure}

\section{Thermodynamic Limit at Fixed Density}\label{sec:thermlim}
Here we introduce a limit which arises in the thermodynamic regime, where both the number of particles \( N \) and the system size \( L \) grow large (\( N, L \to \infty \)) while maintaining a constant particle density \( \rho = N / L \). As illustrated in Fig.\,\ref{fig:comparevst}, the PDF $q_N(t)$ in Eq.\,(\ref{eq:defq}) exhibits features that deviate from both the Gumbel and Fréchet cases, indicating the need for a different framework to describe this intermediate regime.
To derive the extreme value theory for fixed $\rho$, we recap the escape dynamics of a single particle in a one-dimensional box.

\subsection{Distribution of first passage time for a single particle}
We begin by recapping the statistics of the escape of a single particle from a box of size \( L \). The goal is to determine the FPT PDF, which describes the probability of the $i$-th particle escaping the box at time \( t \). The FPT PDF is found by solving the diffusion Eq.\,(\ref{eq:FP}) with an absorbing boundary conditions at \( x = 0 \) and a reflective boundary at \( x = L \), and then spatially integrating. The resulting PDF is expressed as \cite{redner2001guide}
\begin{equation}\label{eq:exactp}
p(t) = \sum_{n=0}^\infty \frac{2 D \pi (n+\frac{1}{2})}{L^2} \sin\left(\pi (n+\frac{1}{2}) \frac{x_0}{L}\right) e^{-D (\pi (n+\frac{1}{2}))^2 t / L^2},
\end{equation}
where \( x_0 \) is the initial position of the particle and $D$ is the diffusion coefficient. 
The case of an infinitely large interval ($L\rightarrow \infty$) has been previously addressed by Schrödinger \cite{schrodinger1915theorie, redner2001guide}, resulting in the PDF given by 
\begin{align}\label{eq:freesol}
    p(t)_\mathrm{free}=\frac{x_0}{\sqrt{4\pi D}} t^{-3/2}\exp\left(-\frac{x_0^2}{4D t}\right) \\ \label{eq:freesol2}\underset{t \gg x_0^2/D}{\sim} \frac{x_0}{\sqrt{4\pi D}}t^{-3/2} \, .
\end{align}
As mentioned above in Eq.\,(\ref{eq:freshtail}), this implies Fr\'echet statistics with $\alpha=1/2$ for the extreme value of the FPT.
\par To account for the boundary effects in a finite system, we consider the limit of large system size and long time. The latter aligns with the principles of extreme value statistics, as it allow us to capture the asymptotic behavior of rare events. For \( x_0 / L \ll 1 \), the argument of the sine function in Eq.\,(\ref{eq:exactp}) remains small, and the PDF simplifies for large \( t \) as
 \begin{align}\label{eq:etasum1d0}
    p(t)\sim 2D\frac{x_0}{L^3}\sum_{n=0}^\infty \pi^2(n+\frac{1}{2})^2 e^{-\pi^2(n+\frac{1}{2})^2Dt/L^2}\, \\ \label{eq:etasum1d}
    = -2\frac{x_0}{L}\partial_{t}\sum_{n=0}^\infty e^{-\pi^2(n+\frac{1}{2})^2Dt/L^2}\, 
\end{align}
In this limit, we define two related scaling functions to encapsulate the large-\( t \) and large-\( L \) behavior. The first, \( \mathcal{I}(\mathcal{\tau}) \), describes the scaled PDF for $x_0/L\rightarrow 0$
\begin{equation}\label{eq:definf}
 p(t)\sim D\,\frac{x_0}{L^{3}}\mathcal{I}(\mathcal{\tau})
\end{equation}
where \( \tau = D \,t / L^2 \) is the rescaled time. Explicitly, from Eq.\,(\ref{eq:etasum1d}), we can write the scaling function as
\begin{equation}\label{eq:isum}
  \mathcal{I}(\mathcal{\tau})  = -2\partial_{\tau}\sum_{n=0}^\infty e^{-\pi^2(n+\frac{1}{2})^2\tau}\, . 
\end{equation}
The second scaling function, \( \mathcal{Z}(\tau) \), relates to $\mathcal{I}(\tau)$ by integration in the same limit as
\begin{equation}
\mathcal{Z}(\mathcal{\tau}) = \int_{\mathcal{\tau}}^\infty d\tau' \, \mathcal{I}(\tau') \, .
\end{equation}
Using Eq.\,(\ref{eq:definf}), we relate $\mathcal{Z}(\tau)$ to the survival probability as
\begin{equation}\label{eq:scal_s_1}
S(t)\sim \frac{x_0}{L}\,\mathcal{Z}(\tau) \,,
\end{equation}
where $S(\cdot)$ is the survival probability of a single particle defined in Eq.\,(\ref{eq:defsurv}). By summing the series in Eq.\,(\ref{eq:isum}), we find \cite{baravi2025solutions}
\begin{equation}\label{eq:inf1first}
 \mathcal{I}(\tau)= -\partial_{\tau }\vartheta_2(e^{-\pi^2 \tau}) \, ,
\end{equation}
where $\vartheta_a(.)$ is the Jacobi elliptic theta function, with $a=2$. 
 Consequently, the statistics of rare, large escape times for a single particle are described by the scaling function
 \begin{align}\label{eq:surv1}
    \mathcal{Z}(\tau)
   = \vartheta_2(e^{-\pi^2 \tau})\, .
\end{align}
Here $\mathcal{Z}(\tau)$ provides a compact form for the single-particle FPT distribution that depends on the system size $L$ only through the rescaled time $\tau$ and is independent of the initial condition $x_0$. 
\par Using the asymptotic of the Jacobi elliptic theta function  $\vartheta_2(\cdot)$, Eq.\,(\ref{eq:surv1}) provides two distinct time dependencies in the limits of $\tau\ll 1$ and $\tau \gg 1$,
\begin{equation}\label{asymp1}
    \mathcal{Z}(\tau)\rightarrow \begin{cases}
 (\pi\, \tau)^{-\frac{1}{2}}  
 , &  \tau\ll 1\\
            2 e^{-\frac{\pi^2}{4} \tau}          , &  \tau \gg 1
                    \end{cases} \, .
\end{equation}
For short times (\( \tau \ll 1 \)), the survival probability exhibits a power-law decay, characteristic of distributions with heavy tails, aligning with the infinite system limit in Eq.\,(\ref{eq:freesol2}); for long times (\( \tau \gg 1 \)), it decays exponentially, consistent with lighter-tailed distributions.

When switching to the EV problem, these asymptotics connect directly to the Gumbel and Fréchet distributions, which emerge as limiting cases within the general solution. Specifically, the exponential decay for \( \tau \gg 1 \) aligns with the Gumbel limit, while the power-law behavior for \(\tau \ll 1 \) reflects the Fréchet limit. However, as we now demonstrate in the thermodynamic limit, the full structure of the scaling function in Eq.~(\ref{eq:surv1}) governs the extreme value statistics, except in the asymptotic regimes of very low or very high density, where limiting forms suffice.

\subsection{Extreme value statistics of the last particle}\label{subsec:ev}
Following this introduction, we now study the statistics of the slowest FPT among a population of $N$ particles escaping from a box of size $L$. The cumulative probability for the largest FPT among \( N \) particles $Q_N(\mathcal{T})$ is given by the expression $(1-S(\mathcal{T}))^N$, which follows from Eq.\,(\ref{eq:nparticles1}). Since extreme-value analysis focuses on rare, long-time events, the asymptotic form of the single-particle survival probability obtained in the previous section becomes central here. In this formulation, the scaling function $\mathcal{Z}(\cdot)$ encapsulates the essential information that governs the extreme-value statistics.
\par In the context of EV, namely in the limit of large $N$, we use the approximation
\begin{equation}\label{eq:firstta}
\left[1 - S(\mathcal{T})\right]^N \sim \left[1 - \frac{x_0}{L} \mathcal{Z}\left(\frac{D\,\mathcal{T}}{L^2}\right)\right]^N=
\left[ 1 - \rho \,\frac{x_0}{N} \mathcal{Z}\left(\frac{D\,\mathcal{T}}{L^2}\right) \right]^N ,
\end{equation}
taking the limits $N,L\rightarrow \infty$, using Eqs.\,(\ref{eq:nparticles1},\ref{eq:firstta}) we get
\begin{equation}\label{eq:qapprox}
\lim_{ N,L \to \infty} Q_N(\mathcal{T}) = \exp{\left[-x_0 \rho\, \mathcal{Z}\left(\frac{D\,\mathcal{T}}{L^2}\right)\right]}.
\end{equation}
where $\rho=N/L$ is the particles density. Note that the transition from Eq.~(\ref{eq:firstta}) to Eq.~(\ref{eq:qapprox}) is justified only when the accompanying exponent $x_{0}\,\rho\,\mathcal{Z}(\tau)$ stays finite as $N\to\infty$. This means that for the rescaled time to remain finite in the limit $L\rightarrow \infty $, the extreme FPT $\mathcal{T}$ must also be large, scaling as $\mathcal{T}\propto L^2$. 
\par
\begin{figure}[h]
  \centering
    \includegraphics[width=0.9\linewidth]{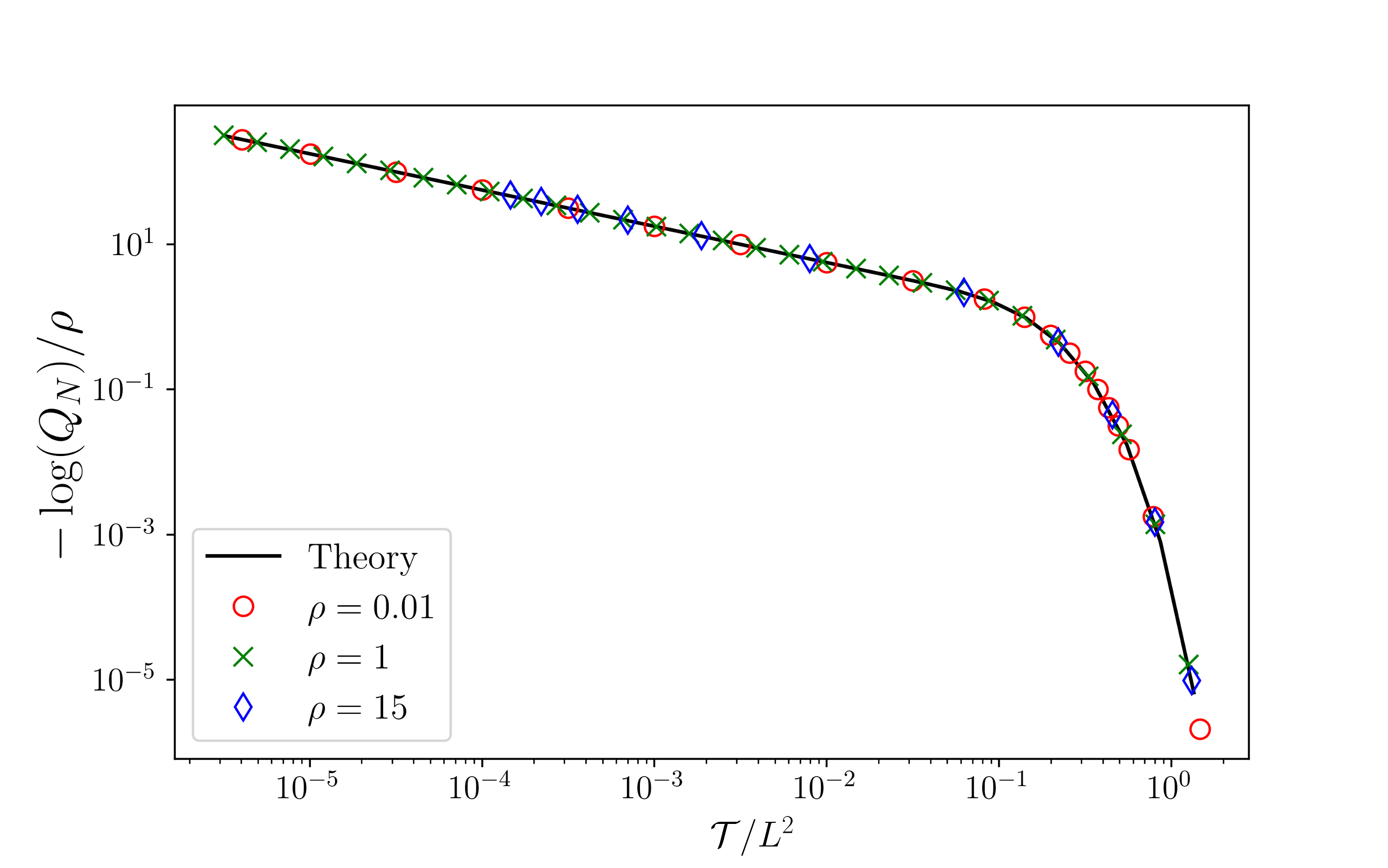}
  \caption{Plot of \(-\log{Q_N} / \rho\) as a function of the rescaled time \( \mathcal{T} / L^2 \) where the number of particles is \( N = 10^4 \). The figure demonstrates the scaling behavior of the cumulative probability for the slowest first-passage time in the thermodynamic limit, where \( \rho = N / L \) remains fixed. The results validate the convergence to the theoretical form described by Eqs.\,(\ref{eq:surv1},\ref{eq:qapprox}). For small $\mathcal{T}/L^2$ we observe a power-law decay consistent with Fr\'echet statistics, while for large values, the distribution exhibits an exponential-like cutoff. The scaling function thus smoothly interpolates between these two regimes. Here we used $D=1/2$ and $x_0=1$.}\label{fig:logqn}
\end{figure}

 To illustrate this result, we plot \(-\log{Q_N}/x_0\rho\) as a function of the rescaled time \( D\,\mathcal{T} / L^2 \) in Fig.\,\ref{fig:logqn}. The results collapse onto a master curve, which is independent of $x_0$ and $\rho$, and is given by \( \mathcal{Z}(\cdot) \) in Eq.\,(\ref{eq:surv1}). Both the figure and the theory demonstrate that in the thermodynamic limit we have an extreme value theory which differs from the standard forms.

\par
\begin{figure}[h]
    \centering
    \begin{minipage}[t]{0.492\linewidth}
        \centering
        \includegraphics[width=\linewidth]{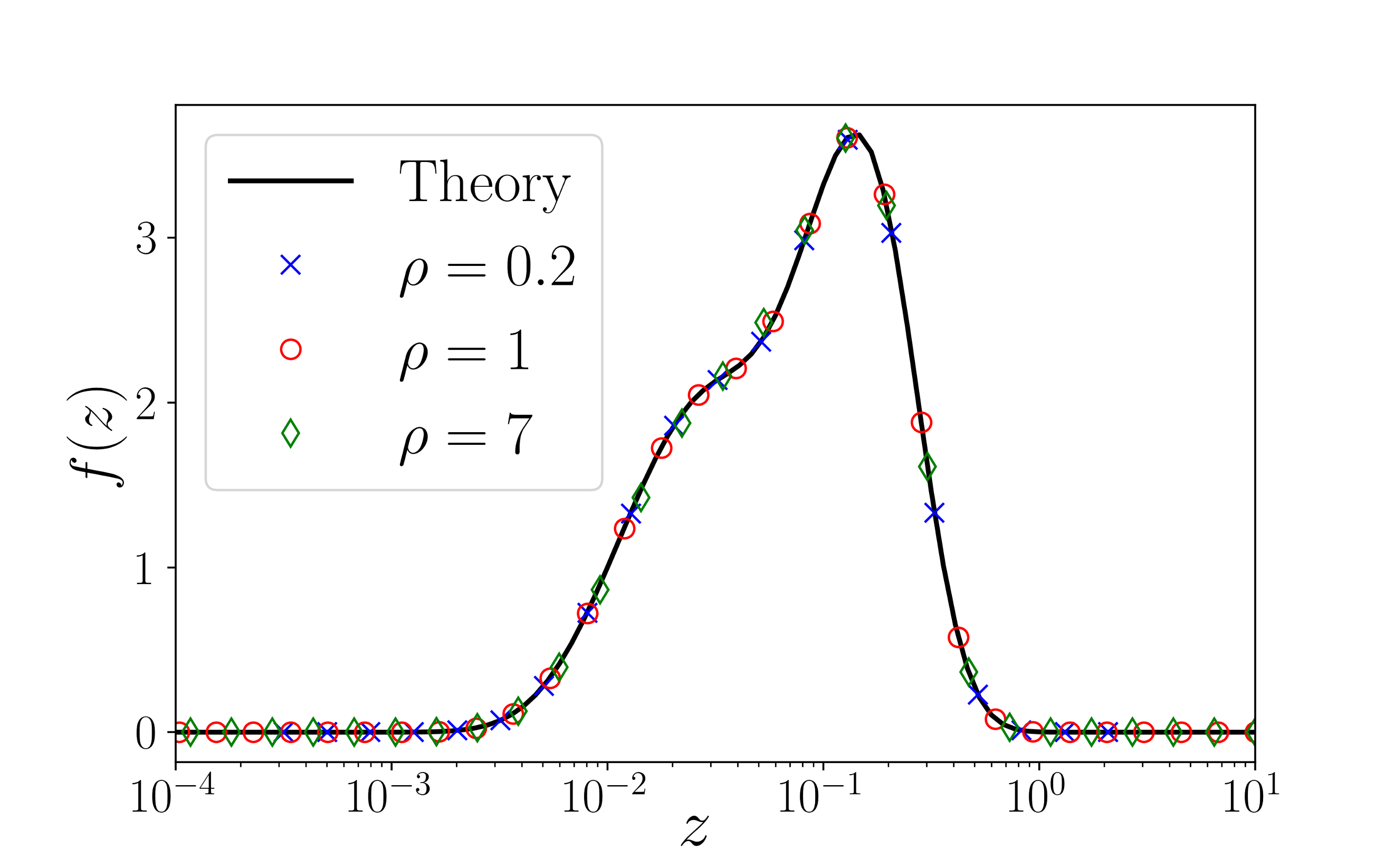}
    \end{minipage}
    \hfill
    \begin{minipage}[t]{0.492\linewidth}
        \centering
        \includegraphics[width=\linewidth]{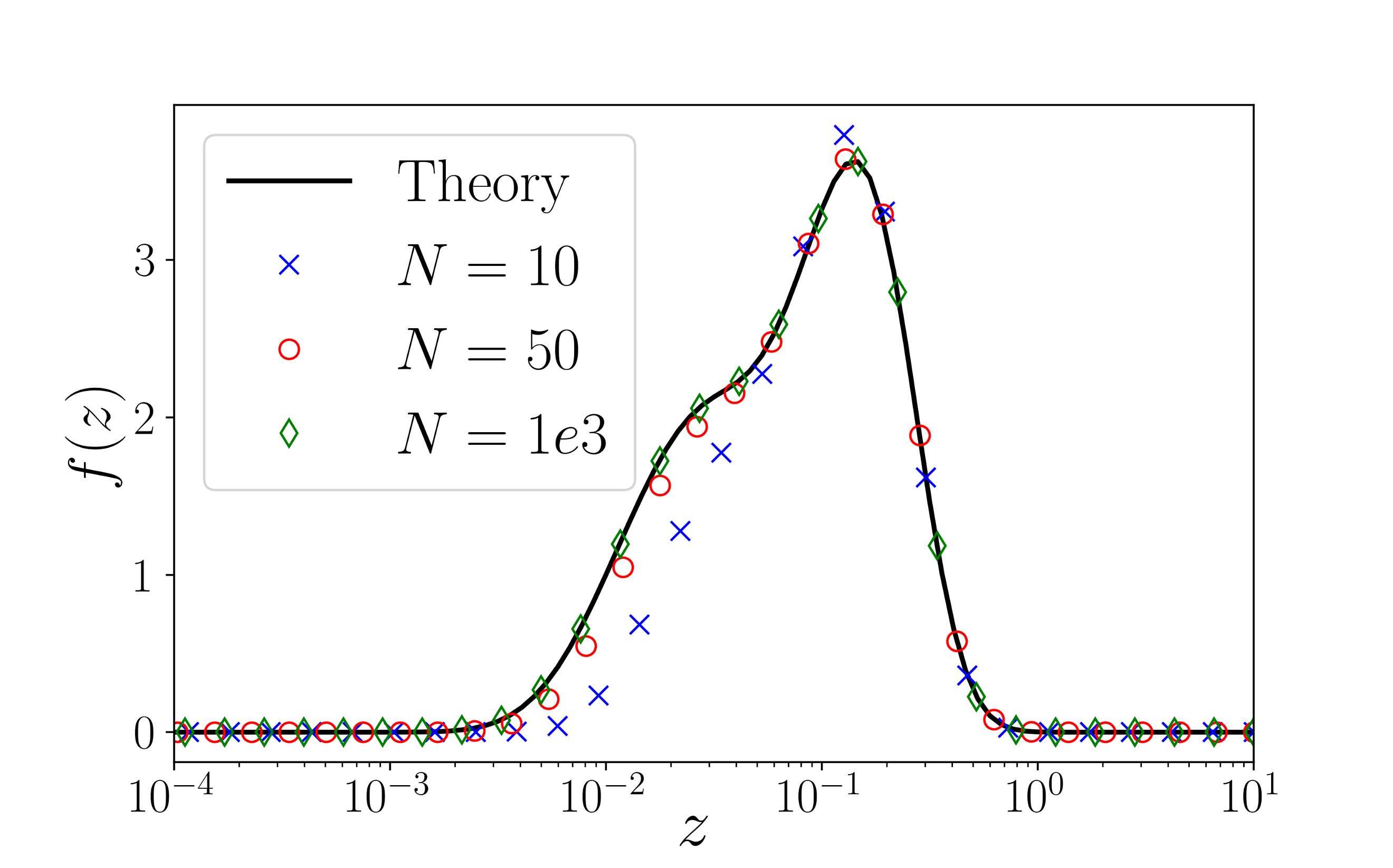}
    \end{minipage}
    \caption{PDF of the slowest first-passage time: (Left) Results for different particle densities \( \rho \) with \( N = 2000 \). The rescaled variable $z$ is defined via Eq.\,(\ref{eq:newzdef}). The solid line represents the theoretical prediction from Eqs.\,(\ref{eq:newzdef}-\ref{eq:theorypdf}), while the symbols denote the exact rescaled solution obtained from Eqs.\,(\ref{eq:defq},\ref{eq:exactp}). (Right) Results for different values of \( N \) at fixed density \( \rho = 1 \), as we increase $N$ keeping $\rho$ fixed we reach the limiting law in Eq.\,(\ref{eq:theorypdf}). The solid black line represents the theoretical prediction from Eqs.\,(\ref{eq:theorys}) and (\ref{eq:theorypdf}), and the symbols show the exact rescaled solution. Here we used \( x_0 = 1 \) and \( D = 1/2 \). The results illustrate the emergence of a new limiting distribution whose shape is distinct from the Gumbel and Fréchet forms shown in Fig.\DIFdelbeginFL \DIFdelFL{,}\DIFdelendFL \DIFaddbeginFL \DIFaddFL{\,}\DIFaddendFL \ref{fig:comparevst}, underscoring the different structure of the finite-density regime.}
    \label{fig:pdfvsz}
\end{figure}

 We now formulate our results in terms of a rescaled variable as the one defined in Eq.\,(\ref{eq:defz1}). Following the limit given in Eq.\,(\ref{eq:qapprox}), the cumulative probability for the slowest FPT approaches a limiting form
\begin{equation}\label{eq:limf}
\lim_{\substack{N,L \to \infty \\ \rho \text{ fixed}}} Q_N(\mathcal{T}) \rightarrow F_{\rho}(z) \, ,
\end{equation}
where the random variable $z$ is related to the random variable $\mathcal{T}$ with the transformation
\begin{equation}\label{eq:newzdef}
\mathcal{Z}(z)=x_0\rho \,\mathcal{Z}\left(\frac{D\,\mathcal{T}}{L^2}\right)\, ,
\end{equation}
and $F_{\rho}(z)$ is given explicitly by
\begin{equation}\label{eq:theorys}
F_{\rho}(z)= \exp\left[-\vartheta_2(e^{-\pi^2 z})\right] \, ,
\end{equation}
following Eq.\,(\ref{eq:surv1}). This formula replaces the Fr´echet and Gumbel limiting laws in Eqs.\,(\ref{eq:gumbcdf},\ref{eq:freshcdf}). The variable \( z \) can be computed semi-analytically using a simple numerical program, see Appendix \ref{appendix:zvst} for details on the numerical procedure and the behavior of \( z \) as a function of \( \mathcal{T} \). From Eq.\,(\ref{eq:theorys}), the PDF for the escape of the slowest particle is given by
\begin{equation}\label{eq:theorypdf}
f_{\rho}(z)= \mathcal{I}(z)\exp\left[-\mathcal{Z}(z)\right] \, . 
\end{equation}
In Fig.\,\ref{fig:pdfvsz}, we plot the exact PDF of the slowest particle, as obtained from Eqs.\,(\ref{eq:defq},\ref{eq:exactp}), alongside the theoretical prediction from Eqs.\,(\ref{eq:newzdef}) and (\ref{eq:theorypdf}). For different fixed values of the particle density \( \rho \), the exact results show excellent agreement with the theoretical predictions, demonstrating the validity of the proposed framework.

\par \DIFaddbegin \DIFadd{The introduction of the rescaled variable $z$ in Eq.~(\ref{eq:newzdef}) serves the same purpose as the classical transformation used in Eq.~(\ref{eq:defz1}): it removes the system-dependent parameters ($x_0,L,\rho$) and allows the distributions at different densities to collapse onto a single universal reduced form, as illustrated in Fig.~\ref{fig:pdfvsz}.
In this sense, Eq.~(\ref{eq:newzdef}) plays the role of the classical change of variables that underlies the Gumbel, Fréchet, and Weibull limits in standard extreme-value theory, while Eq.~(\ref{eq:theorypdf}) provides the analogue of the limiting functions given in Eqs.~(\ref{eq:gumbelpdf}) and~(\ref{eq:frechetpdf}). 
This construction places our theory within the conventional extreme-value framework, enabling comparison between the slowest-first-passage distribution derived here and the classical limiting laws.
 }

\par \DIFaddend We can now revisit the limiting cases presented in Section \ref{sec:second}. The excellent agreement between our theoretical predictions and the exact numerical results in Figs.\,\ref{fig:comparegumbel}–\ref{fig:comparevst} is a direct consequence of this framework. In particular, Figs.\,\ref{fig:comparegumbel} and \ref{fig:comparefresh} illustrate the convergence to the Gumbel and Fréchet laws, given in Eqs.\,(\ref{eq:gumbelpdf}) and (\ref{eq:frechetpdf}), while simultaneously showing how our theory in Eqs.\,(\ref{eq:newzdef},\ref{eq:theorys}) captures the same data across densities. Fig.\,\ref{fig:comparevst} further highlights that neither classical limit suffices on its own, and that the full scaling description is needed Since our theory offers a unified framework across different density values, it naturally includes the Gumbel and Fréchet distributions as limiting cases. We now demonstrate how these classical extreme value statistics arise under specific conditions.

\subsection{ High-density Limit }\label{subsec:gumbel}
Since $\mathcal{Z}(\tau)$ in Eq.\,(\ref{eq:surv1}) is monotonically decreasing in $\tau$, finiteness of $x_{0}\,\rho\,\mathcal{Z}(\tau)$ implies a direct correspondence between $\rho$ and $\tau$. For fixed $x_0$, lowering $\rho$ must be compensated by a smaller $\tau$ (short times), whereas increasing $\rho$ pushes the relevant $\tau$ to larger values. Therefore, to study the high-density limit, we examine the behavior of the scaling function \(\mathcal{Z}(\cdot)\) in the large argument regime. From Eq.\,(\ref{asymp1}) we found $\mathcal{Z}(D\,\mathcal{T}/L^2) \sim 2 e^{-\frac{\pi^2}{4} D\,\mathcal{T}/L^2}$, for  $D\,\mathcal{T}/L^2 \gg 1$, which demonstrates that the survival probability decays exponentially for large extreme FPT as expected for a bounded system. This is consistent with the conditions for the Gumbel distribution, which arises for exponential-tailed parent distributions. 
From Eq.\,(\ref{eq:qapprox}), the cumulative probability \( Q_N(\mathcal{T}) \) in this regime is approximated as
\begin{equation}
Q_N(\mathcal{T}) \sim \exp{\left[-x_0 \rho \, \mathcal{Z}\left(\frac{D\,\mathcal{T}}{L^2}\right)\right]}\, .
\end{equation}
Substituting with the asymptotic form of \(\mathcal{Z}(\cdot)\) we obtain
\begin{equation}\label{eq:qngumbel}
Q_N(\mathcal{T}) \sim \exp\left[-x_0 \rho \cdot 2 e^{-\frac{\pi^2}{4} \frac{D\,\mathcal{T}}{L^2}}\right].
\end{equation} 
The exponential decay of \( Q_N(\mathcal{T}) \) matches the Gumbel form when
   \begin{equation}
   2x_0\rho\,e^{-\frac{\pi^2}{4} \frac{D\,\mathcal{T}}{L^2}} \sim e^{-z}.
   \end{equation}
   From this, the rescaled variable in Eq.\,(\ref{eq:newzdef}) becomes
\begin{equation}\label{eq:abgumbel}
z \rightarrow \frac{4\,\mathcal{T}}{\pi^2\,L^2}  -  \log 2x_0\rho   \, .
\end{equation}
This logarithmic scaling of the shift parameter with $\rho$ (and thus with $N$) is consistent with the behavior expected from the Gumbel distribution \cite{majumdar2020extreme}. We see the same compensation between density and time:
to keep \(z\) finite while \(\mathcal{T}/L^{2}\to\infty\), the factor \(x_{0}\rho\) must
grow correspondingly in logarithmic rate, therefore this is indeed a high-density limit. This shows that our general theory correctly reduces to the Gumbel distribution in the high-density limit where \( N \gg L / x_0 \).

\subsection{ Low-density Limit }
Let us now consider the behavior of \(\mathcal{Z}(\cdot)\) in the short-time regime, corresponding to low density. From Eq.\,(\ref{asymp1}), we find $\mathcal{Z}(D\,\mathcal{T}/L^2) \sim (\pi D\,\mathcal{T}/L^2)^{-\frac{1}{2}}$ for $D\,\mathcal{T}/L^2 \ll 1$. Using Eq.\,(\ref{eq:scal_s_1}), the survival probability exhibits a power-law decay for intermediate rescaled time, namely $x_0^2/L^2\ll \tau \ll 1 $. This power-law decay is consistent with the conditions for the Fréchet distribution, which arises for heavy-tailed parent distributions.
Substituting the asymptotic form of \(\mathcal{Z}(\cdot)\) into the cumulative probability function in Eq.\,(\ref{eq:qapprox}), we find
\begin{equation}\label{eq:qnfreshet}
Q_N(\mathcal{T}) \sim \exp\left[-x_0 \rho \, \left(\pi \frac{D \,\mathcal{T}}{L^2}\right)^{-\frac{1}{2}}\right].
\end{equation}
To match this with the standard Fréchet distribution in Eq.\,(\ref{eq:freshcdf}), we identify the rescaled variable \( z \) through the transformation in Eq.\,(\ref{eq:newzdef}). The power-law decay of \( Q_N(\mathcal{T}) \) matches the Fréchet form when
\begin{equation}
x_0\rho \left(\pi \frac{D \,\mathcal{T}}{L^2}\right)^{-\frac{1}{2}} \sim z^{-\alpha}.
\end{equation}
Therefore, for the Fréchet distribution with a tail index \(\alpha = 1/2\), we identify
\begin{equation}\label{eq:abfrechet}
z \rightarrow \frac{D\,\pi\,\mathcal{T} }{x_0^2 N^2} \, ,
\end{equation}
which matches the results expected in the limit $L\rightarrow \infty$ (see Appendix \ref{appendix:Gumbel}). 
Note that since this limit is taken for $L\rightarrow \infty$ for finite time so the assumption in Eq.\,(\ref{eq:limf}) is for $t/L^2 \rightarrow 0$. To keep $z$ finite we have to assume finite $N$ such that $N/L=\rho\rightarrow 0$, thus this is a low-density limit.
Thus, in the limit of short \( t \), the survival function \(\mathcal{Z}(\cdot)\) exhibits a power-law tail, leading to Fréchet statistics for the slowest FPT. This demonstrates how our general theory reduces to the Fréchet distribution in the low-density limit, where \( N \ll L / x_0 \).

\DIFaddbegin \subsection{\DIFadd{Localized and Random Initial Conditions}}
\DIFadd{So far, we have focused on the regime $x_0/L \to 0$, where the scaling function $\mathcal{I}(\cdot)$, characterizing the rare events of the parent distribution, provides a new formulation of the extreme-value problem. In Appendix C, we examine the complementary case $x_0 = \alpha L$. The comparison reveals how the asymptotic theory performs for finite $\alpha$: for $\alpha \lesssim 0.1$, the agreement remains satisfactory. This indicates that as long as $\alpha$ is small, the theory based on the infinite invariant density offers a robust approximation, showing only weak sensitivity to variations in $\alpha$.
}

\DIFadd{The theory developed in this work applies to a localized initial state $x_0$. It is also natural to consider cases where $x_0$ is a random variable. In Appendix \ref{appendix:initial}, we study the corresponding extreme-value problem for initial positions uniformly distributed in the interval $x_0\in[0,L]$, considering both quenched and annealed formulations. These provide a complementary perspective to the localized setup discussed above.
}

\DIFaddend \section{The moments of the slowest FPT}\label{sec:extmom}
We now study the moments of the EVs. These show different behaviors that crucially depend on the order of the moment. Low-order moments and high-order moments differ in their dependence on the system size $L$. These distinctions are crucial for characterizing the full statistical properties of the slowest FPT, especially in the intermediate regimes that bridge classical limits.
\subsection{Mean extreme first-passage time }

To further characterize the statistics of the slowest particle, we consider the first moment of its first-passage time, defined as
\begin{equation}\label{eq:meandef}
    \langle \mathcal{T}\rangle = \int_0^{\infty} q_N(\mathcal{T}') \, \mathcal{T}' \, d\mathcal{T}',
\end{equation}
where as mentioned $q_N(\mathcal{T})$ is the PDF of the slowest first-passage time among $N$ particles in a system of length $L$. From Eqs.\,(\ref{eq:defq}) and (\ref{eq:qapprox}) in the limit $L,N\rightarrow \infty$ where $\rho=N/L$ is maintained constant, $q_N(\mathcal{T})$ is given explicitly as
\begin{equation}\DIFaddbegin \label{eq:momqnremind}
    \DIFaddend q_N(\mathcal{T})\approx x_0 \rho\, \frac{\partial}{\partial \mathcal{T}}\exp{\left[-x_0 \rho \, \mathcal{Z}\left(\frac{D\,\mathcal{T}}{L^2}\right)\right]}\, ),
\end{equation}
 where $Z(\cdot)$ is the scaling function derived in Eq.\,(\ref{asymp1}). The corresponding expression for the mean extreme first-passage time (MEFPT) is
\begin{align}\label{eq:defmean1}
\langle \mathcal{T} \rangle &\approx \DIFaddbegin \DIFadd{-}\DIFaddend \frac{L^2}{D} x_0 \rho \int_0^\infty \tau\, \vartheta_2'(e^{-\pi^2 \tau}) \exp\left[-x_0 \rho\, \vartheta_2(e^{-\pi^2 \tau})\right] d\tau\, ,
\end{align}
where $\tau = D t / L^2$. Using the definition of the Jacobi elliptic theta function $ \vartheta_2(e^{-\pi^2 t})=2\sum_{n=0}^\infty e^{-\pi^2(n+1/2)^2 t}$, we can rewrite Eq.\,(\ref{eq:defmean1}) as
\begin{align}\label{eq:mfpt1}
\langle \mathcal{T} \rangle & \approx \frac{L^2}{D}2 x_0 \rho \sum_{n=0}^\infty \frac{1}{\pi^2(n+1/2)^2} \int_0^\infty y\, e^{-y-2x_0\rho e^{-y}}  R_n(y)dy 
\end{align}
where 
\begin{align}\label{eq:defrn}
R_n(y) &\equiv \prod_{m\neq n}^\infty\exp{\left[-2x_0 \rho\, e^{-\lambda_{n,m}\,y}\right]} \, ,
\end{align}
and $\lambda_{n,m}=(m+1/2)^2/(n+1/2)^2$. \DIFaddbegin \DIFadd{Derivation details are provided in Appendix~\ref{app:mefpt}. }\DIFaddend This expression is general and holds for all values of the particle density $\rho$. Numerical evaluation of Eq.\,(\ref{eq:mfpt1}) produces results that match the exact calculation with excellent accuracy across the entire density range as we soon demonstrate.

Let us study the two limits of the MEFPT for low and for high density $\rho $. In the high-density limit, the system is expected to be governed by Gumbel statistics, as discussed previously in Section~\ref{subsec:gumbel}. In this limit, the integrand $e^{-y-2x_0\rho e^{-y}}$ in Eq.\,(\ref{eq:mfpt1}) becomes sharply peaked around $y\sim \log{2x_0 \rho}$. Expanding around $y=\log{2x_0\rho} + \Delta y$ we rewrite Eq.\,(\ref{eq:defrn}) as
\begin{align}\label{eq:limrn}
R_n(y) &\approx\prod_{m\neq n}^\infty\exp{\left[-(2x_0 \rho)^{1-\lambda_{n,m}\,}\, e^{-\lambda_{n,m}\,\Delta y}\right]} \,.
\end{align}
The product in Eq.\,(\ref{eq:limrn}) tends to zero when $\lambda_{n,m}<1$ (i.e., for $n>m$), and to one when $\lambda_{n,m}>1$ (i.e., for $m>n$). The only index $n$ for which the condition $\lambda_{n,m}>1$ holds for all $m\neq n$ is $n=0$. Consequently, the dominant contribution in the sum over $n$ arises solely from the $n=0$ term. This leaves us with
\begin{equation}\label{eq:rnisdelta}
    R_n(y)\approx\delta_{n,0}\, .
\end{equation}
Substituting Eq.\,(\ref{eq:rnisdelta}) into Eq.\,(\ref{eq:mfpt1}), we obtain
\begin{align}\DIFaddbegin \label{eq:midexpangumb}
\DIFaddend \langle \mathcal{T} \rangle &\approx\frac{L^2}{D}2 x_0 \rho \sum_{n=0}^\infty \frac{1}{\pi^2(n+1/2)^2} \delta_{n,0}\int_0^\infty y\, e^{-y-2x_0\rho e^{-y}}  dy \\
&=\frac{8 x_0 \rho \,L^2}{\pi^2D}  \int_0^\infty y\, e^{-y-2x_0\rho e^{-y}}  dy \, , 
\end{align}
which is nothing but the first moment of the Gumbel distribution in Eq.\,(\ref{eq:gumbelpdf}), given by
\begin{equation}\label{eq:meangumb}
\langle \mathcal{T} \rangle \approx \frac{4\,L^2}{\pi^2 D} \left( \gamma + \Gamma(0,2 x_0 \rho) + \log(2 x_0 \rho) \right),
\end{equation}
where $\gamma \approx 0.577$ is Euler’s constant and 
$\Gamma(0, z) = \int_{z}^{\infty} t^{-1} e^{-t}\, dt$
denotes the incomplete Gamma function. This result is known in the context of the Gumbel statistics \cite{lawley2023slowest}, as expected in the high-density limit where the exponential decay of the survival probability dominates.

In the low-density limit, $x_0\,\rho \ll 1$, we can replace the summation over $m$ in Eq.\,(\ref{eq:defrn}) with an integral over $k_{n,m}\equiv \sqrt{\lambda_{n,m}}$, namely
\begin{align}
R_n(y) &=\exp{\left[-2x_0 \rho\,\sum_{m\neq n} e^{-k_{n,m}^2\,y}\right]} \\ & \approx \exp{\left[-2x_0 \rho\,\left(\int_0^\infty dk \,e^{-k^2\,y} -e^{-y}  \right)\right]} \\&\approx\exp{\left[-2x_0 \rho\,\left(\frac{1}{2}\sqrt{\frac{\pi}{y}}-e^{-y}\right)\right]}\, .
\end{align}
Plugging this into Eq.\,(\ref{eq:mfpt1}) yields 
\begin{align}\label{eq:mfptlowrho}
\langle \mathcal{T} \rangle
&\approx\frac{L^2}{D}2 x_0 \rho \sum_{n=0}^\infty \frac{1}{\pi^2(n+1/2)^2} \int_0^\infty y\, e^{-y-2x_0\rho e^{-y}}  R_n(y)dy\\& \approx \frac{L^2}{D}2 x_0 \rho \sum_{n=0}^\infty \frac{1}{\pi^2(n+1/2)^2} \int_0^\infty y\, e^{-y-x_0\rho \sqrt{\pi/y}}  dy \,.
\end{align}
Using the identity $\sum_{n=0}^\infty \frac{1}{\pi^2(n+1/2)^2}=1/2$ and performing the integration over $y$, we find
\begin{equation}\label{eq:lowdenmfpt}
     \langle \mathcal{T} \rangle\approx \frac{L^2}{D} x_0 \rho \frac{1}{\sqrt{\pi}} \, G^{0,0}_{3,0} \left( \frac{(x_0\rho) ^2 \pi}{4} \,\middle|\, \begin{matrix} 0,\; \tfrac{1}{2},\; 2 \\ \end{matrix} \right)\, .
\end{equation}
where $G(\cdot)$ is the Meijer G-function. Using the small-argument asymptotics of the Meijer G-function $G^{0,0}_{3,0} \left( \frac{x^2 \pi}{4} \,\middle|\, \begin{matrix} 0,\; \tfrac{1}{2},\; 2 \\ \end{matrix} \right)\sim \sqrt{\pi}$, Eq.\,(\ref{eq:lowdenmfpt}) scales as \begin{equation}\label{eq:mefptlowrho}
    \langle \mathcal{T} \rangle \rightarrow \frac{\rho\,L^2 x_0}{D}=N \langle t \rangle_{SP}\, ,
    \end{equation}
namely a linear dependence on the number of particles multiplied by the scaling of the MFPT of a single particle $\langle t \rangle_{SP}=x_0L/D$. 
    Notably, the explicit low-density expression for the MEFPT derived from our theoretical framework represents a distinct asymptotic result, different from the predictions of both Gumbel and Fréchet statistics. Although the slowest FPT distribution appears Fréchet-like at low densities, the associated MFPT for a true Fréchet law diverges, which is non-physical for any finite-sized system. This divergence is evident in Eq.\,(\ref{eq:mfpt1}) when considering the naive expansion of our theory in the $\rho \rightarrow 0$ limit. In contrast, our result remains finite and well-defined for any finite $\rho$.

\begin{figure}[h]
    \centering
     \includegraphics[width=0.8\linewidth]{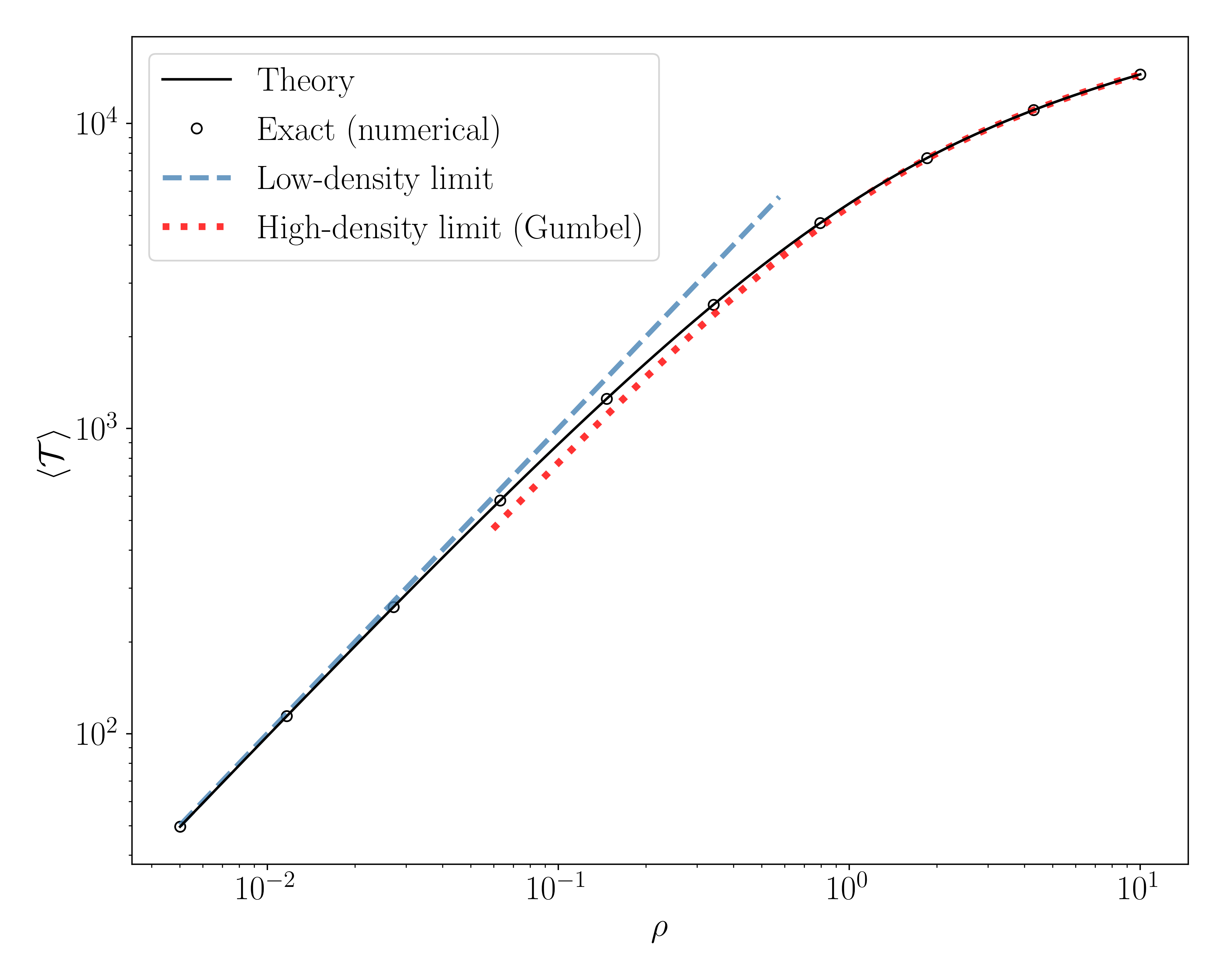}
    \caption{ MEFPT as a function of the density $\rho$ on a log-log scale for fixed $L=100$ and $x_0=1$. The exact results (black symbols) are obtained via numerical integration of Eqs.\,(\ref{eq:defq},\ref{eq:exactp}). The theoretical prediction in Eq.\,(\ref{eq:defmean1}) (solid black line) agrees with the exact result across all $\rho$. The low-density approximation from Eq.\,(\ref{eq:mefptlowrho}) (dashed blue line) accurately captures the behavior at small $\rho$, while the high-density limit from Eq.\,(\ref{eq:meangumb}) (dotted red line) describes the large-$\rho$ regime, consistent with predictions based on Gumbel statistics.  Although the general statistics of the maximum FPTs at low density are governed by Fr\'echet laws, this framework alone predicts a diverging mean time as $\rho \rightarrow 0$, therefore, our theoretical approach is necessary to correctly describe the MEFPT in the low-density limit.
}
    \label{fig:mommfpt}
\end{figure}

As shown in Fig.\,\ref{fig:mommfpt}, the Gumbel approximation for the MEFPT is accurate only in the high-density limit and breaks down for intermediate values of $\rho$. This is the regime our theory addresses. We compare the exact MFPT (obtained from direct numerical integration of Eqs.\,(\ref{eq:defq}) and (\ref{eq:exactp}) with the prediction from our theory using Eq.\,(\ref{eq:defmean1}), with the Gumbel approximation in Eq.\,(\ref{eq:meangumb}) and with the low-density limit in Eq.\,(\ref{eq:mefptlowrho}).
The results are shown for fixed $L=1000$, as a function of the density $\rho$. While the Gumbel-like prediction becomes accurate only for larger $\rho$, our theory agrees with the exact result across the entire range.

\subsection{The q-moment}
\begin{figure}[h]
    \centering
    \includegraphics[width=0.99\linewidth]{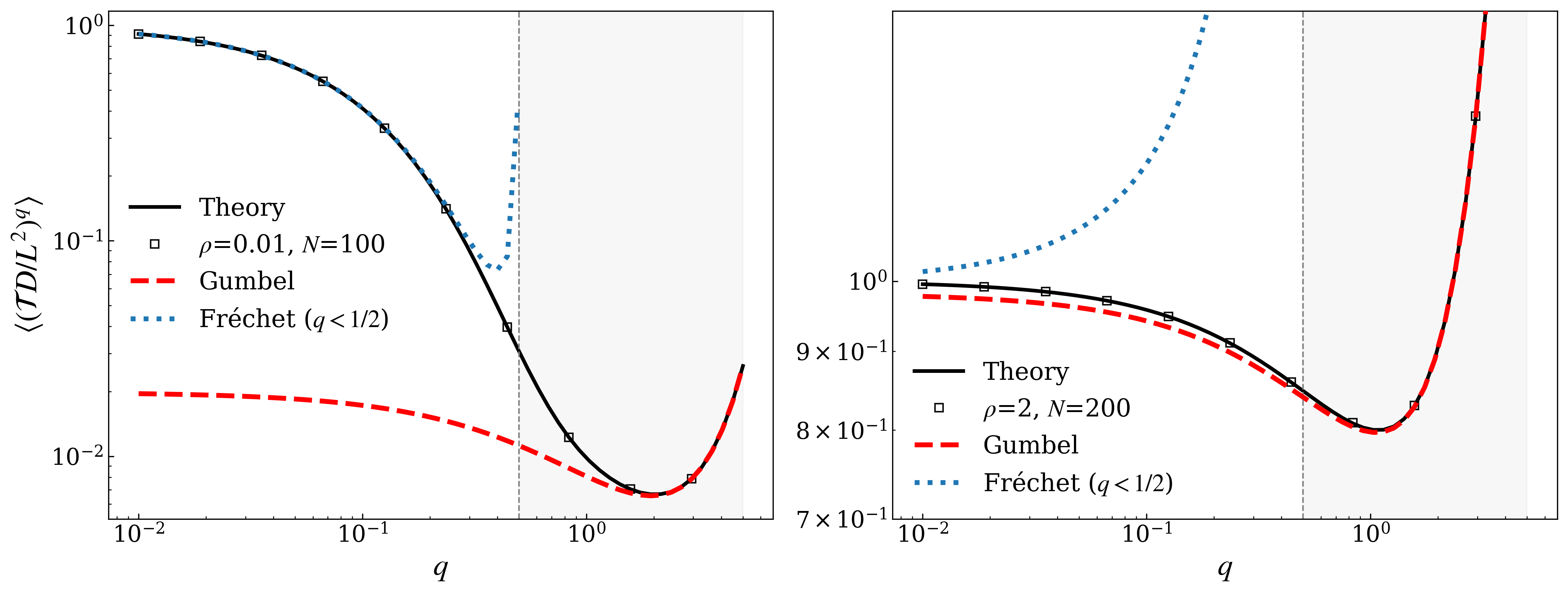}
    \caption{Dimensionless moments of slowest first-passage time, $\langle\mathcal{ (D\,T)}^{q}\rangle/L^{2q}$, as a function of the moment order $q$ for $\rho=0.01$ (left) and $\rho=2$ (right). We have set $x_0=1$ and $D=1/2$. The gray dashed vertical line marks the transition point at $q=1/2$. Symbols represent exact numerical calculations, while the solid black line shows our theoretical prediction in Eq.\,(\ref{eq:genqmom}). The panels illustrate the crossover between extreme-value regimes. At low density (left panel), our theory agrees with the Fréchet prediction for low-order moments ($q < 1/2$) in Eq.\,(\ref{eq:momfresh}), but captures the finite moments for all $q$, whereas a naive application of Fréchet law would predict a divergence as $q \to 1/2$. Conversely, at high density (right panel), our theory and the Gumbel prediction in Eq.\,(\ref{eq:momgumb}) overlap. As shown, neither classical limit suffices, while the distinctive nature of our theory bridges the Fréchet-like and Gumbel-like behaviors.}
    \label{fig:momqn}
\end{figure}

We now consider the general $q$-th moment of the slowest first-passage time $\mathcal{T}$, defined by
\begin{equation}
\langle \mathcal{T}^q \rangle = \int_0^{\infty} q_N(t) \, t^q \, dt , \qquad q\ge 0 \, .
\end{equation}
From Eqs.\,(\ref{eq:surv1},\ref{eq:qapprox}) the asymptotics of the q-moment can be written as 
\begin{equation}\label{eq:genqmom}
\langle \mathcal{T}^q \rangle \approx  - x_0 \rho \int_0^\infty t^q\, \vartheta_2'(e^{-\pi^2 D\, t/L^2}) \exp\left[-x_0 \rho\, \vartheta_2(e^{-\pi^2 D\, t/L^2})\right] dt \, . \end{equation}
In the high-density limit (\( x_0\,\rho \gg 1 \)), as we soon show, the moments converge to those of the Gumbel distribution (see Fig.\,\ref{fig:momqn}), reflecting the dominance of the exponential decay in the parent survival probability. Nevertheless, for finite densities, the behavior of the moments is more nuanced. For low-order moments (\( q < 1/2 \)), the convergence is to the moments of the Fréchet distribution, capturing the influence of the heavy-tailed nature of the distribution in this regime. However, for high-order moments (\( q > 1/2 \)), the behavior is no longer captured by the classical limits. In this regime, the moments converge to the predictions of our finite-density theory, which deviates from both the Gumbel and Fréchet forms.
\par To illustrate this in the low-density regime, we consider the asymptotic form of Eq.\,(\ref{eq:genqmom}) in the short-time limit. Using the asymptotics of the Elliptic theta function $\vartheta_2(e^{-\pi^2 x})\sim (\pi \, x)^{-1/2}$ for small $x$, we obtain
\begin{equation}\label{eq:momqfrechet}
\langle \mathcal{T}^q \rangle_{\rho \ll 1} \sim  \frac{x_0 N}{\sqrt{4\pi D}} \int_0^\infty t^{q-3/2}\exp\left[-x_0 N\, (\pi D\, t)^{-1/2}\right] dt \,. \end{equation}
It is easy to see that Eq.\,(\ref{eq:momqfrechet}) diverges for $q>1/2$, including the MFPT. For $q<1/2$ the integral gives
 \begin{equation}\label{eq:momfresh}
 \langle \mathcal{T}^q \rangle_{\text{Fr\'echet},q<\frac{1}{2}} = \Gamma(1-2q) \left(\frac{x_0N}{\sqrt{\pi D}}\right)^{2q}\, . \end{equation}

In contrast, in the high-density limit (\( x_0\,\rho \gg 1 \)), the distribution of the slowest first-passage time becomes sharply peaked and exhibits Gumbel-like behavior. For sufficiently high densities and moments \( q > 1/2 \), the moments are expected to converge to those of the Gumbel distribution. Using the asymptotics of the Elliptic theta function $\vartheta_2(e^{-x})\sim 2e^{-x/4}$ for large $x$, Eq.\,(\ref{eq:genqmom}) reduces to 
\begin{equation}\label{eq:gumbgen}
\langle \mathcal{T}^q \rangle_{\rho \gg 1} \sim   2 x_0 \rho \frac{\pi^2 D}{4L^2} \int_0^\infty t^q\, e^{-\pi^2 D\, t/L^2} \exp\left[-2x_0 \rho\,e^{-\pi^2 D\, t/4L^2}\right] dt \, . \end{equation}
While a closed-form solution of Eq.\,(\ref{eq:gumbgen}) for a general value of $q$ is difficult to obtain, the case of non-negative integer moments \( \langle \mathcal{T}^n \rangle \) can be computed explicitly. Simplify Eq.\,(\ref{eq:gumbgen}), the \(n\)th moment becomes
\begin{align}
\langle \mathcal{T}^n \rangle_{\text{Gumbel}} &= 2 x_0 \rho \left(\frac{4}{\pi^2}\right)^n \left(\frac{L^2}{D}\right)^n \int_0^\infty y^n \exp\left[-y - 2\,x_0 \rho\, e^{-y}\right] dy,\\ \label{eq:momgumb}&
 = 2 x_0 \rho \left(\frac{4}{\pi^2}\right)^n \left(\frac{L^2}{D}\right)^n n! \cdot \,_nF_n\left(\underbrace{1,\dots,1}_{n+1}; \underbrace{2,\dots,2}_{n+1}; -2x_0 \rho \right).
\end{align}
where \( \,_nF_n(a_1,\dots,a_n; b_1,\dots,b_n; z) \) is the generalized hypergeometric function.

As illustrated in Fig.\,\ref{fig:momqn}, neither the Gumbel nor the Fréchet moments alone fully capture the behavior of the moments across all densities or orders $q$. In contrast, our theoretical framework, based on the numerical evaluation of Eq.\,(\ref{eq:genqmom}) accurately reproduces the full range of moment behaviors. It provides an excellent fit to the exact numerical results for all values of $\rho$ and $q$. This highlights the limitations of classical Gumbel and Fr\'echet statistics in describing the full behavior of the system. The crossover behavior, especially for intermediate densities, requires a more complete description, precisely what is provided by our finite-density theory.

\section{Extension to compact diffusion on a fractal domain}\label{sec:exten}
Next we discuss how the theory developed in this work can be extended beyond the example of particle in a one-dimensional box to a wide class of confined compact stochastic processes. A stochastic process is said to be compact if a single particle, in the absence of boundaries, returns to the origin with probability one.
This recurrence property holds whenever the walk dimension $d_w$ exceeds the fractal dimension $d_f$ of the substrate on which the walk takes place, i.e., when $d_w > d_f$. The marginal case, $d_f=d_w$, while considered recurrent, behaves more like the non-compact regime \cite{meyer2011universality}. Therefore, in this work, we focus exclusively on the strictly compact case $d_w>d_f$. The fractal dimension $d_f$ of the medium quantifies how the mass $M(R)$ (e.g., number of sites in a cluster) within radius $R$ scales with $R$, namely $M(R)\sim R^{d_f}$, and thus how densely the substrate fills space across scales \cite{ben2000diffusion,gefen1983anomalous,rammal1983random}. The walk dimension $d_w$ quantifies the temporal spread of the process in the absence of absorption, through the scaling of the mean-square displacement $\langle r^2(t) \rangle \sim K\,t^{2/d_w}$, with $K$ been a generalized diffusion constant \cite{havlin2002diffusion,bouchaud1990anomalous} \cite{Note1}. A third relevant exponent is the the persistence exponent $\theta$ that describes the long-time decay of the survival probability in unbounded systems as $S(t) \propto t^{-\theta}$ \cite{majumdar1999persistence}.

\begin{figure}
    \centering
    \includegraphics[width=0.4\linewidth]{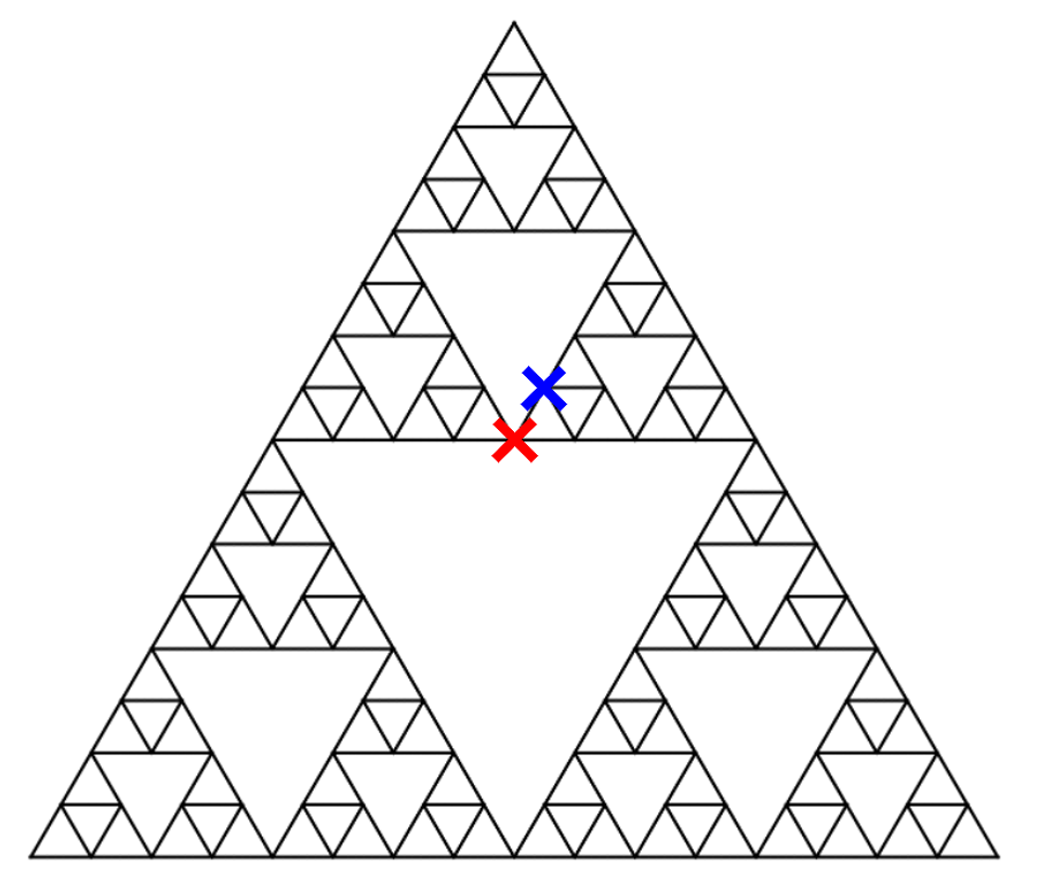}
    \caption{Schematics of a discrete Sierpiński gasket with a depth of 5-iterations; the red vertex indicates the initial position, while the blue vertex is the absorbing site, here $r_0=1$ is the chemical distance from the initial position to the target. Note that in this example the source and target are single lattice sites, so the geometry is not radially symmetric as in the O–P continuum model. The specific value of the initial condition $r_0$ enters through the distribution given in Eqs.\,(\ref{eq:newzdef2}) and (\ref{eq:generalqn}). The walkers are performing a discrete-time random walk on this graph. The fractal and walk dimensions of the gasket are $d_f = \ln 3 / \ln 2$ and $d_w = \ln 5 / \ln 2$, respectively. The confining length $L$ is defined via the total number of sites $n$ by $L = n^{1/d_f}$ \cite{meyer2011universality}. On the Sierpi´nski gasket, the number of nodes at iteration (generation) $m$ is given by $n=(3+3^{m+1})/2$ \cite{teguia2005sierpi}. Here for $m=5$ the confining length is $L\approx 41.4$ and the generalized diffusion coefficient is $K\sim 0.365$. For discussion of first-passage times for single particles on this structure, see \cite{baravi2025first,meyer2011universality}.}
    \label{fig:schematics}
\end{figure}

Here we consider the FPT of a diffusing particle that starts at a fixed distance $r_0$ from an absorbing target, with the condition that the confining length scale $L$ is much larger than $r_0$, i.e., $L \gg r_0$, see Fig.\,(\ref{fig:schematics}). The maximal FPT is then defined as the time taken for the last particle among $N$ independent walkers to reach the absorbing target.

\par Our following analysis builds upon the general framework developed in our earlier work \cite{baravi2025solutions}, which relies on the universality approach of Meyer et al.\,\cite{meyer2011universality}. In \cite{baravi2025solutions} we derived scaling solutions for a broad class of compact first-passage processes in confined domains. There, we showed that under the compact exploration condition $d_w > d_f$, the survival probability of a single particle in a confined system admits a universal structure which can be expressed in the general scaling form
\begin{equation}\label{eq:gens}
    S(t) \sim \left( \frac{r_0}{L} \right)^{d_w \theta} \mathcal{Z}\left( \tau \right), \qquad \tau = \frac{t}{\tau_L}\, ,
\end{equation}
where $\tau_L=(L/K)^{d_w}$ and the generalized diffusion constant $K$ is defined below. This is a direct generalization of Eq.~(\ref{eq:scal_s_1}) which was derived for the case of Brownian motion in a 1D box. In \cite{baravi2025solutions} the reader will find the explicit forms of the scaling function $\mathcal{Z}(\cdot)$ for various widely applicable models. The functional form of $\mathcal{Z}(\cdot)$ depends on the underlying dynamics and geometry, but it is independent of $L$ and $r_0$. 
Consequently, the distribution of the largest FPT out of $N$ independent walkers is given by

\begin{equation}\label{eq:generalqn}
    Q_N(\mathcal{T}) \xrightarrow[\rho,\;\mathcal{T}/\tau_L\;\text{ fixed}]{\mathcal{T},\, N,\, L \to \infty} 
    \exp\left[ -r_0^{d_w \theta} \rho^*\, \mathcal{Z}\left( \frac{ \mathcal{T}}{\tau_L} \right) \right],
\end{equation}
where the effective density $\rho^*$ is defined as
\begin{equation}\label{eq:effden}
    \rho^* \equiv \frac{N}{L^{d_w \theta}}.
\end{equation}
It is important to recognize that this thermodynamic limit involves three diverging quantities ($\mathcal{T},N,L \rightarrow \infty$ keeping $\mathcal{T}/\tau_L$ and $\rho^*$ finite), compared to the two-variable limits that yield the classical Gumbel and Fréchet laws. Note that while $\rho^*$ plays the role of an effective particle density in our theory, it does not always carry the physical units of a conventional density (i.e., particles per unit volume). This is because the exponent $d_w \theta $ generally differs from the spatial dimension $d_f$, and thus $L^{d_w\theta}$ does not represent a geometric volume. Alternatively, the product $r_0^{d_w\theta}\rho^*$ appearing in the exponent of Eq.\,(\ref{eq:generalqn}) can be interpreted as a dimensionless control parameter. Consequently, the fixed-density limit implies that $N$ scales as $(L / r_0)^{d_w \theta}$.

To connect our generalized framework with the classical results, we can reformulate the problem in terms of a rescaled variable $z$, in the spirit of the Fréchet and Gumbel laws shown in Eqs.\,(\ref{eq:gumbcdf},\ref{eq:freshcdf}).
Following the definition in Eq.~(\ref{eq:defz1}), we introduce a rescaled variable $z$ that absorbs both $L$ and the density $\rho^*$. It is defined implicitly by equating the two arguments of the scaling function $\mathcal{Z}$, viz. 
\begin{equation}\label{eq:newzdef2}
    \mathcal{Z}(z)\;=\;r_0^{d_w\theta}\rho^*\,\mathcal{Z}\!\left(\frac{\mathcal{T}}{\tau_L}\right)\, .
\end{equation}
This yields a unique transformation from the random variable representing the extreme FPT $\mathcal{T}$ and $z$. Expressed in terms of $z$, the extreme–value distribution takes the compact universal form
\begin{equation}\label{eq:f(z)}
    F(z)= \exp\left[-\mathcal{Z}(z)\right]\, .
\end{equation}
All model‐specific information is thus encoded into the single scaling function $\mathcal{Z}$, while the dependence on system size and particle number enters only through the rescaled variable $z$.

In the previous sections, we demonstrated that the distribution of the maximal FPT undergoes a crossover between different extreme-value universality classes as the particle density $\rho^*$ is varied. At low densities, where particles are sparse and the maximal FPT is governed by the intermediate-time behavior of the single-particle survival probability, the distribution follows Fréchet statistics. As higher densities, more particles probe longer times, eventually reaching the regime where the survival probability is exponentially suppressed due to confinement, resulting in a Gumbel distribution. Importantly, the nature of this transition is controlled by the asymptotic behavior of the scaling function $\mathcal{Z}(z)$, which depends on the underlying dynamics and confinement mechanism. As we now demonstrate, diffusion on a fractal geometry provides a concrete realization of the transition from Fréchet to Gumbel statistics, with the crossover controlled by the exponents $d_w$ and $d_f$ that determine the form of $\mathcal{Z}(z)$

Following the general framework presented above, consider the single-particle scaling function $\mathcal{Z}(\tau)$, defined from the survival probability, exhibits a short-time power-law $\mathcal{Z}(\tau)\propto \tau^{-\theta}$ for $\tau\ll 1$. The persistence exponent $\theta =1-d_f/d_w$ fixes the first extreme-value limit: at sufficiently low density $\rho^*$ the maximum FPT follows a Fréchet distribution with tail index~$\theta$. 
As we shall see below,  at longer times or higher densities, the statistics cross over to a Gumbel regime controlled by the long-time cutoff of the parent FPT distribution. In what follows we focus on compact diffusion on a fractal domain, exemplified by the Sierpiński gasket; see Fig.\,\ref{fig:schematics}.

Following the general framework presented above, consider the case of compact random walks on a fractal structure that satisfy the compact exploration condition $d_w > d_f$, such as the Sierpiński gasket. The propagator $P(r,t)$ for a particle diffusing in a fractal domain follows the O'Shaughnessy–Procaccia equation \cite{o1985analytical,o1985diffusion}\cite{Note2},
\begin{equation}\label{eq:diff_frac}
\frac{\partial}{\partial t}P(r,t)=\frac{K}{r^{d_f-1}}\frac{\partial}{\partial r}\left( r^{1+d_f-d_w}\frac{\partial}{\partial r}P(r,t) \right),
\end{equation}
where the normalization condition, determined by the fractal dimension $d_f$ is\DIFaddbegin \DIFadd{, in the absence of absorption
}\DIFaddend \begin{equation}
\int_0^\infty P(r,t)\,r^{d_f-1}dr=1 \, .
\end{equation}
\DIFaddbegin \DIFadd{When an absorbing boundary at $r=0$ is introduced, probability is no longer conserved and the
integral equals the survival probability $S(t)<1$ at time $t$. }\DIFaddend The initial condition is given by $P(r,0) =r_0^{1-d_f} \delta(r - r_0)$, with an absorbing boundary at $r=0$, and a reflecting boundary at $r=L$ $\left. \frac{\partial}{\partial r}P(r,t) \right|_{r = L} = 0$. In Eq.\,(\ref{eq:diff_frac}), $K$ is a generalized diffusion coefficient, and the exponents $d_f$ and $d_w$ are the fractal and walk dimensions, respectively.

\par A canonical physical realization of such a system is the random walk on a Sierpiński gasket. For this specific structure, the exponents are known to be $d_f = \ln 3 / \ln 2$ and $d_w = \ln 5 / \ln 2$. The confining length scale $L$ in the continuum model is related here to the discrete construction of the gasket, for instance, through the number of generations or total number of sites $n$ \cite{meyer2011universality}.
From Eq.\,(\ref{eq:effden}), the effective density on fractals follows directly from 
\begin{equation}
\rho^* \;=\;\frac{N}{L^{d_w \theta}}
\quad\text{with}\quad
\theta = 1 - \frac{d_f}{d_w}
\;\;\Longrightarrow\;\;
\rho^* \;=\;\frac{N}{L^{\,d_w - d_f}}.
\end{equation}
As mentioned before, $\rho^*$ does not carry the units of a conventional spatial density (which would be $N/L^{d_f}$). Thus, $\rho^*$ should be viewed as an effective density.

\par The scaling functions $\mathcal I(\tau),\mathcal{Z}(\tau)$ has been derived in \cite{meyer2011universality,baravi2025solutions}. It is obtained from the eigenmode expansion of the O'Shaughnessy–Procaccia operator in Eq.\,(\ref{eq:diff_frac}) and takes the form
\begin{align}
  \mathcal I(\tau) &=
     \frac{2^{2\nu-3}\,d_w}{\Gamma(2-\nu)}
     \sum_{n=0}^{\infty}
        \frac{J_\nu(z_{-\nu,n})}{J_{1-\nu}(z_{-\nu,n})}\,
        z_{-\nu,n}^{\,3-2\nu}\,
        e^{-\,d_w^{2}z_{-\nu,n}^{2}\tau/4}, \qquad  \nu \equiv \frac{d_f}{d_w}<1
     \label{eq:I_fractal}\\[4pt]
  \mathcal Z(\tau) &= \int_{\tau}^{\infty}\!\mathcal I(u)\,du= \frac{2^{2\nu-1}}{d_w\,\Gamma(2-\nu)}
     \sum_{n=0}^{\infty}
        \frac{J_\nu(z_{-\nu,n})}{J_{1-\nu}(z_{-\nu,n})}\,
        z_{-\nu,n}^{\,1-2\nu}\,
        e^{-\,d_w^{2}z_{-\nu,n}^{2}\tau/4}
     ,
     \label{eq:Z_fractall}
\end{align}
where $z_{-\nu,n}$ denotes the $n$-th positive zero of the Bessel function $J_{-\nu}(z)$. Here again, $\tau = t/\tau_L$ with $\tau_L = L^{d_w}/K$, where $K$ in Eq.\,(\ref{eq:diff_frac}) is defined through the mean-square displacement via $\langle r^2(t)\rangle=\frac{\Gamma((2+d_f)/d_w)}{\Gamma(d_f/d_w)}(d_w^2 K)^{2/d_w}\, t^{2/d_w}$.
\par
\begin{figure}[h]
    \centering
    \begin{minipage}[t]{0.49\linewidth}
        \centering
        \includegraphics[width=\linewidth]{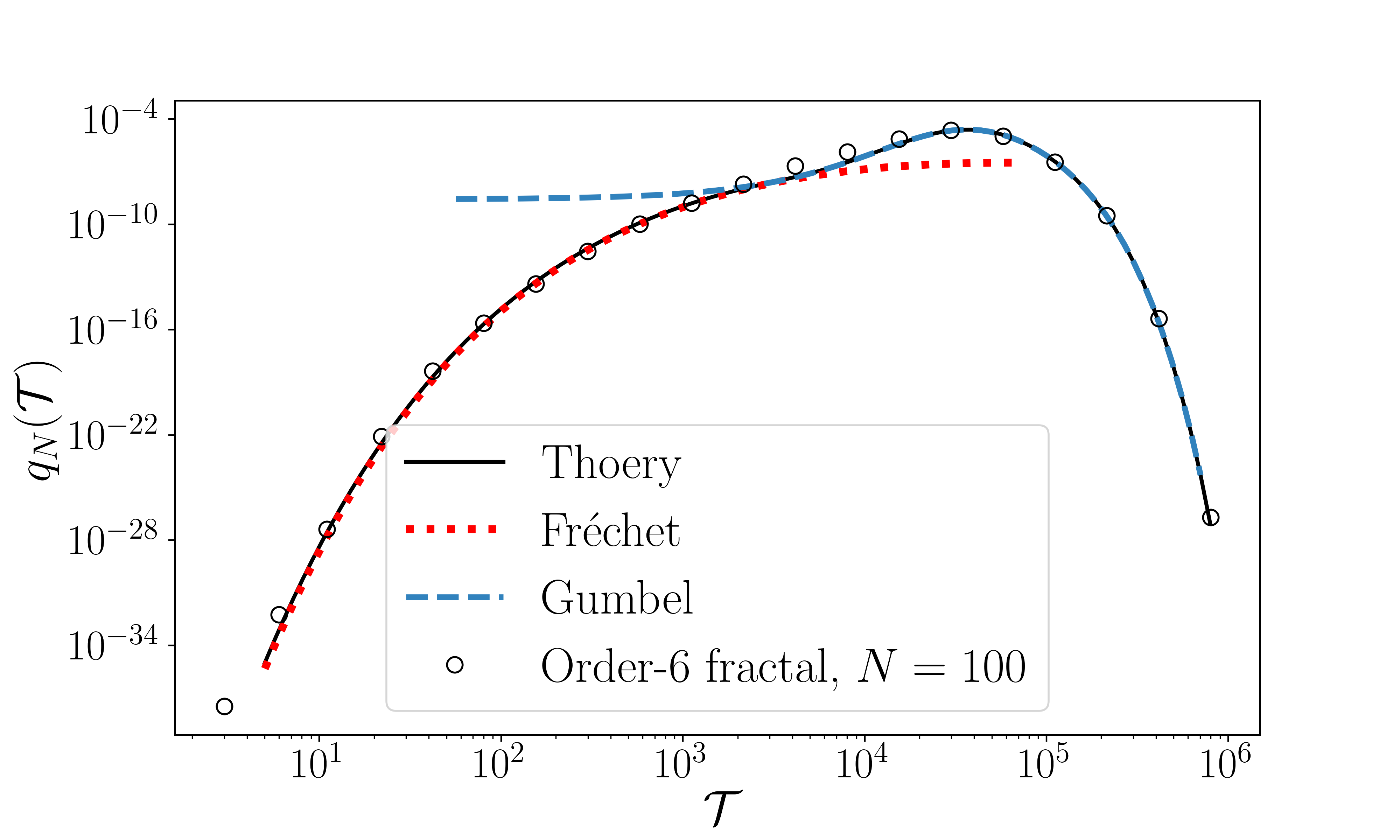}
    \caption{
Extreme–value crossover for diffusion on a fractal domain with $d_f=\log{3}/\log{2}$ and $d_w=\log{5}/\log{2}$. The solid black curve represents the theoretical PDF of the slowest FPT, $q_N(\mathcal{T})$, obtained from Eqs.\,(\ref{eq:generalqn},\,\ref{eq:I_fractal},\,\ref{eq:Z_fractall}) at effective density $\rho^\ast = N/L^{\,d_w-d_f} = 3.8$. The symbols are obtained from direct numerical solution to the master equation of random walks on a discrete Sierpinski gasket of order-6\DIFaddbeginFL \DIFaddFL{, see appendix \ref{appendix:simulation} for details}\DIFaddendFL . The dotted line is the Fréchet limit with tail index $\theta = 1 - d_f/d_w$ in Eq.\,(\ref{eq:zlimfrac}), valid when $\mathcal{T}$ is small; the dashed line shows the Gumbel distribution, which dominates for large~$\mathcal{T}$. }
\label{fig:fractal2}
    \end{minipage}
    \hfill
    \begin{minipage}[t]{0.49\linewidth}
        \centering
\includegraphics[width=\linewidth]{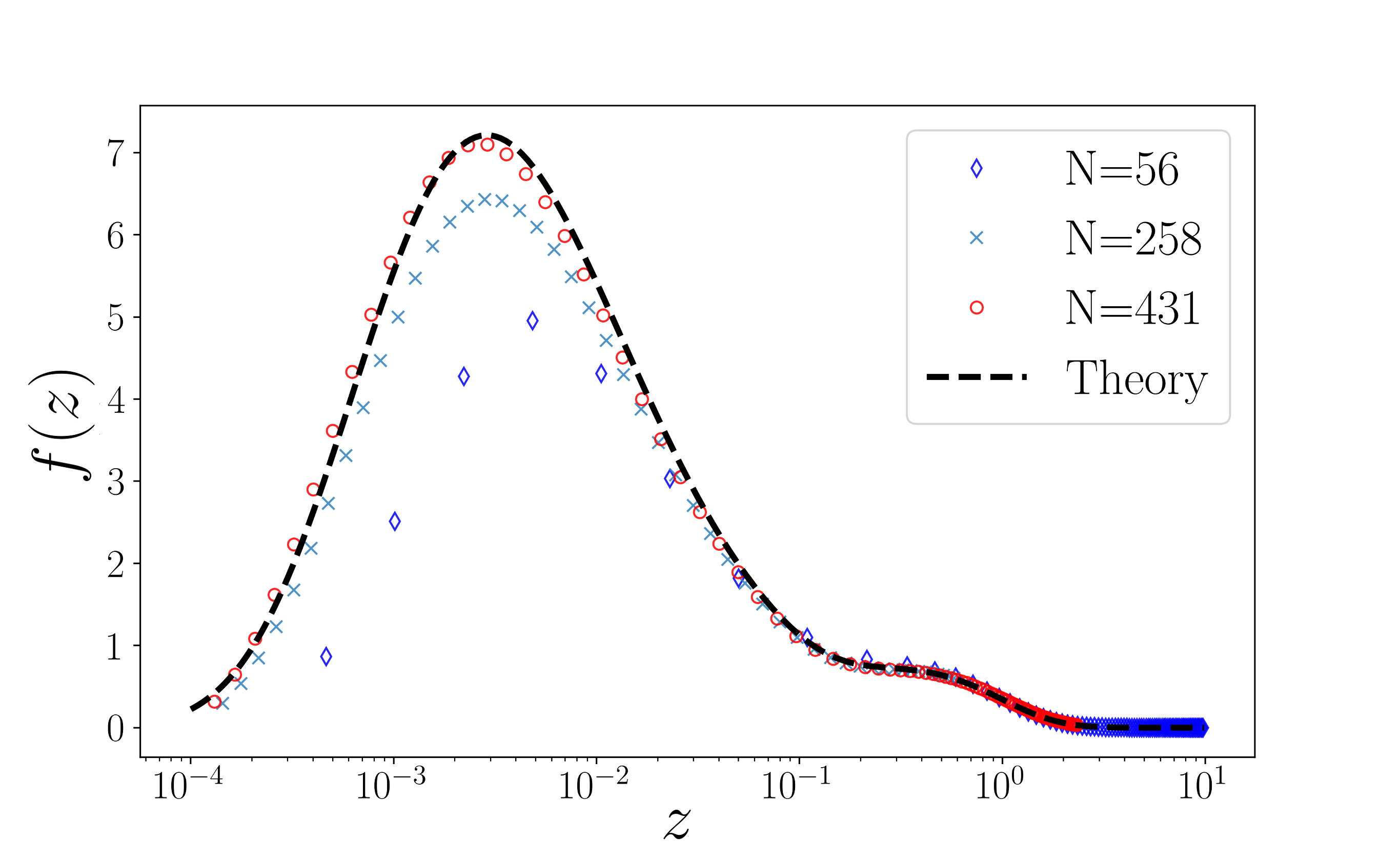}
        \caption{Convergence of the rescaled extreme–value PDF to the scaling function $f(z)$ for diffusion on a fractal geometry.
        The dashed black line shows the theoretical prediction derived from the continuum O’Shaughnessy–Procaccia diffusion model using Eqs.\,(\ref{eq:theorypdf},\ref{eq:I_fractal},\ref{eq:Z_fractall}), with $d_f=\log{3}/\log{2}$ and $d_w=\log{5}/\log{2}$. The rescaled variable $z$ is obtained via the mapping in Eq.\,(\ref{eq:newzdef2}). The symbols represent data obtained from direct numerical simulations of random walks on a discrete Sierpiński gasket\DIFaddbeginFL \DIFaddFL{, see appendix \ref{appendix:simulation} for details}\DIFaddendFL . Simulation details are provided in Appendix \ref{appendix:simulation}. All data sets correspond to the same effective density $\rho^* = N / L^{d_w-d_f} = 10$. As the number of particles $N$ increases, the simulation results collapse onto the universal theoretical curve.
}\label{fig:fractal}
    \end{minipage}
\end{figure}

 The asymptotic behavior of $\mathcal{Z}(\tau)$ governs the limiting form of the maximal first-passage time distribution. It exhibits the following behavior in the short- and long-time regimes as
\begin{equation}\label{eq:zlimfrac}
   \mathcal Z(\tau)\;\rightarrow\;
   \begin{cases}
      \frac{d_w^{-2+2\nu} }{2(\nu-1) \Gamma(\nu) } \tau^{-1+\nu}, & \tau\ll 1,\\[4pt]
      c_0\exp\!\bigl[-\lambda_0(d_f,d_w)\,\tau\bigr], & \tau\gg 1,
   \end{cases}
\end{equation}
where
$\displaystyle 
   \lambda_0 = d_w^{2}\,z^{2}_{-d_f/d_w,0}/4
$ and $c_0=(2^{2\nu-1}\,J_\nu(z_{-\nu,0})\,z_{-\nu,0}^{\,1-2\nu})/(d_w\Gamma(2-\nu)\,J_{1-\nu}(z_{-\nu,0}))$.
By substituting the two limiting forms of $\mathcal{Z}(\tau)$ into the general expression for the distribution of the maximum FPT in Eqs.\,(\ref{eq:generalqn},\ref{eq:f(z)}) we obtain 
\begin{itemize}[leftmargin=1.2em]
\item \emph{Fréchet:}
      for low density, namely $r_0^{d_w \theta}\rho^*\ll 1$, which as shown in section \ref{subsec:ev} corresponds to the short-time limit,  
      $F(z)\sim
      \exp\!\Bigl[-\,c_1\,z^{-(1-\nu)}\Bigr]$ with $c_1=d_w^{-2+2\nu} /(2(\nu-1) \Gamma(\nu) )$
     . Using Eq.\,(\ref{eq:newzdef2}) we find $z=r_0^{-d_w}(\mathcal{T}/\tau_L)(\rho^*)^{-1/(1-\nu)}$.

\item \emph{Gumbel:}
      for high density $\rho^*$,  
      $F(z)\sim
      \exp\!\Bigl[-c_0\,
                  \exp(-\lambda_0 z)\Bigr]$, where, from Eq.\,(\ref{eq:newzdef2}) we find $z=\mathcal{T}/\tau_L-\log{(r_0^{d_w-d_f}\rho^*)}/\lambda_0$.
\end{itemize}
Hence, diffusion on a fractal domain exhibits a clear Fréchet-to-Gumbel crossover in the distribution of the maximal first-passage time, controlled by the effective density $\rho^*$ and the time scale $\tau_L$. Fig.~\ref{fig:fractal2} shows this crossover for the Sierpiński gasket, illustrating the two asymptotic regimes as $\rho^*$ is varied. Importantly, our general theoretical framework, expressed in Eqs.\,(\ref{eq:newzdef2},\ref{eq:f(z)},\ref{eq:Z_fractall}), captures both limiting behaviors, Fréchet at short extreme times and Gumbel at long extreme times, within a single expression.

\par This example underscores that, while the formulas in Eqs.\,(\ref{eq:generalqn},\ref{eq:newzdef2},\ref{eq:f(z)}) remains universal, the functional form of $\mathcal Z(\tau)$, depends sensitively on the geometry and the specific values of $d_f,\,d_w$. The convergence of the rescaled maximum first-passage time distribution to $F(z)$ is demonstrated in Fig.~\ref{fig:fractal}, confirming the predictive power of the framework developed in this section.

\section{Conclusions}
At the first stage of this work, we have presented a detailed study for the statistics of the slowest extreme FPT among \( N \) independent Brownian particles confined in a one-dimensional box. We focused on the thermodynamic limit, where both the number of particles \( N \) and the system size \( L \) tend to infinity while maintaining a fixed density \( \rho \). This framework captures and interpolates between two classical universal regimes of the EV statistics: Gumbel statistics for high-density systems with exponential tails, and Fréchet statistics for low-density systems with heavy-tailed power-law distributions. Importantly, our theory also accounts for the intermediate regime where neither classical distributions alone suffices, providing a complete picture of the slowest FPT distribution. The resulting theory is confirmed numerically, showing excellent agreement over a broad range of particle numbers and densities.
\par Although Gumbel and Fréchet statistics are widely regarded as universal, our work shows that the limiting shape of the extreme value distribution in the thermodynamic limit depends on the specific model. What remains general is not the limiting distribution itself, but rather the scaling structure: given knowledge of the corresponding scaling solution \( Z(\tau) \) defined in Eqs.\,(\ref{eq:gens},\ref{eq:newzdef2}), one can construct the theory.
The scaling function \( Z(\tau) \) describes the long-time statistics of the FPT problem for a single particle; in that context, it describes rare events \cite{baravi2025first,baravi2025solutions}. However, in the many-particle scenario, this same function becomes a key and necessary ingredient of the theory. This generality enables a direct extension to other classes of stochastic processes. In particular, as shown in \cite{baravi2025solutions}, in a wide range of systems, the long-time behavior of a single particle distribution is governed by an infinite density function, corresponds to the solution in Eq.\,(\ref{eq:inf1first}). Therefore, the precise form of the scaling function \( Z(\tau) \), and the definition of the effective density \( \rho^* \) will vary with the geometry or dynamics, but the underlying structure of our approach remains valid. In this way, the theory developed here serves as a versatile tool for characterizing extreme first-passage times of $N$ random walkers in a broad class of systems.

Our results establish a complete scaling theory for the statistics of the slowest FPT in confined stochastic systems. While the scaling form of the distribution is universal and expressed through a rescaled variable $z$ and a cumulative function $F(z) = \exp{[-Z(z)]}$, the scaling function $Z(\tau)$ itself depends on the underlying dynamics and geometry, encoded in the exponents $d_w$ and $d_f$. This non-universality reflects the fact that different compact processes, generate distinct extreme-value crossovers. Once $Z(\tau)$ is determined from the single-particle problem (by theory or measurement), all statistical properties of the extreme FPTs follow. This places slowest first-passage problems within extreme-value theory and identifies a distinct limit law specific to compact dynamics in confinement.

\section{Acknowledgments}
We would like to thank to the Israel Science Foundation for their support through grant number 1614/21.

\newpage
\appendix
\section*{Appendices}
\section{Fr\'echet and Gumbel statistics}\label{appendix:Gumbel}

This appendix provides explicit expressions for the classical extreme-value limits discussed in Sec.~II and recovered as limiting cases of our finite-density theory. In particular, we summarize the rescaling parameters and limiting forms of the slowest first-passage time (FPT) distribution in the Gumbel and Fréchet regimes \cite{lawley2023slowest}.

\subsection{A.1 Gumbel Limit: Finite Domain, Large $N$}\label{append1a}

In bounded domains, the first-passage time distribution of a single particle exhibits exponential decay at long times. Accordingly, the maximum of $N$ i.i.d.\ variables with such a parent distribution converges to the Gumbel law as $N \to \infty$.

Let $p(t)$ denote the single-particle FPT distribution, and define the cumulative distribution function for the maximum as
\begin{equation}
    Q_N(\mathcal{T}) = \left[1 - \int_{\mathcal{T}}^\infty p(t')\, dt'\right]^N = [1 - S(\mathcal{T})]^N.
\end{equation}
To obtain a nontrivial limiting form as $N \to \infty$, we rescale time as
\begin{equation}
    z = \frac{\mathcal{T} - a_N}{b_N},
\end{equation}
where $a_N$ and $b_N$ are chosen to match the asymptotic tail of the survival probability.

For Brownian motion in a one-dimensional interval of length $L$ with $D = 1/2$ and $x_0 \ll L$, the long-time behavior of the survival probability is
\begin{equation}
    S(\mathcal{T}) \sim 2 \frac{x_0}{L} \exp\left[-\frac{\pi^2 D \,\mathcal{T}}{4L^2}\right].
\end{equation}
Setting $x_0 = 1$ and matching to the standard Gumbel form $\exp(-e^{-z})$, we find the rescaling parameters
\begin{align}
    a_N &= \frac{4L^2}{\pi^2 D} \log(2 \rho), \\
    b_N &= \frac{4L^2}{\pi^2 D},
\end{align}
with $\rho = N/L$ the particle density. This leads to
\begin{equation}\label{eq:aq6}
    Q_N(\mathcal{T}) \to \exp\left[-e^{-z}\right], \quad z = \frac{\mathcal{T} - a_N}{b_N}.
\end{equation}
The corresponding probability density function is
\begin{equation}
    q_N(\mathcal{T}) \to \frac{1}{b_N} \exp\left[-z - e^{-z}\right].
\end{equation}
The mean value of $T$ in this regime is given by the expression \cite{lawley2023slowest}
\begin{equation}\label{eqappend:gumbav}
    \langle \mathcal{T} \rangle \sim a_N + \gamma b_N = \frac{4L^2}{\pi^2 D} \left( \log(2 \rho) + \gamma \right),
\end{equation}
where $\gamma \approx 0.577$ is Euler’s constant. We use Eq.\,(\ref{eqappend:gumbav}) in Fig.\,\ref{fig:mommfpt}; the data reveal clear deviations from this in the thermodynamic limit at fixed finite density, $\rho \equiv N/L^{d_f}$.

\subsection{A.2 Fréchet Limit: Unbounded Domain, Fixed $N$}\label{append1b}

In unbounded domains, the FPT distribution typically has a heavy tail. For a free Brownian particle starting at $x_0 > 0$ on the half-line, the FPT to the origin is given in Eq.\,(\ref{eq:freesol}) \cite{schrodinger1915theorie}
\begin{equation}
    p(t) = \frac{x_0}{\sqrt{4\pi D t^3}} \exp\left[-\frac{x_0^2}{4Dt}\right],
\end{equation}
which behaves as $t^{-3/2}$ at long times. Consequently, the maximum of $N$ such variables converges to the Fréchet distribution with tail index $\alpha = 1/2$:
\begin{equation}
    Q_N(\mathcal{T}) \to \exp\left[-z^{-\alpha}\right], \quad z = \frac{\mathcal{T}}{a_N}, \quad a_N \sim N^2.
\end{equation}
Equivalently, for the slowest FPT among $N$ particles,
\begin{equation}
    q_N(\mathcal{T}) \sim \frac{x_0 N}{\sqrt{4\pi D \mathcal{T}^3}} \exp\left[-\frac{x_0 N}{\sqrt{\pi D \,\mathcal{T}}}\right],
\end{equation}
which matches the Fréchet form:
\begin{equation}
    f(z) \sim \alpha z^{-(1+\alpha)} \exp\left[-z^{-\alpha}\right], \quad \alpha = \frac{1}{2}.
\end{equation}
Note that for $\alpha \leq 1$, the mean diverges. This is consistent with the divergence of $\langle T \rangle$ in the pure Fréchet limit, in contrast to the finite mean predicted by our thermodynamic-limit theory at small but nonzero density.

\section{ Numerical Solution for new scaled variable $z$ and its limits}\label{appendix:zvst}

We reuse the scaling variable $z$ defined in Eqs\,(\ref{eq:limf},\ref{eq:newzdef}). Here we record the short- and long-$z$ asymptotics and the identities that connect the finite-size (Gumbel) and infinite-domain (Fréchet) limits. The scaled variable \( z \) in our theory is determined by solving the equation:
\begin{equation}
\mathcal{Z}(z)=x_0\rho \,\mathcal{Z}\left(\frac{D\,\mathcal{T}}{L^2}\right)\,,
\end{equation}
where \( \mathcal{Z}(\cdot) \) is a scaling functions defined in terms of the survival probability of the slowest first-passage time. Here, \( z \) depends on the density \( \rho = N / L \) and the rescaled time \(D\,\mathcal{T}/L^2 \). Let us use $x_0,D=1$ for simplification.

To compute \( z \) as a function of \(t\), we solve this equation numerically using Python's \texttt{fsolve} function from the \texttt{scipy.optimize} module, which is designed to find roots of non-linear equations. The process involves:

1. \textbf{Input Functions:} Define the functions \(\mathcal{Z}(z)\) and \(\mathcal{Z}(t)\) based on their respective analytical forms. For a particle in a box, using Eq.\,(\ref{eq:surv1})
   \begin{equation}
   \mathcal{Z}(z) = \vartheta_2(e^{-\pi^2 z}),
   \end{equation}
   where \(\vartheta_2\) is the Jacobi theta function.

2. \textbf{Equation Setup:} The root-finding problem is set up as:
   \begin{equation}
   f(z) \equiv \mathcal{Z}(z) - \rho \mathcal{Z}(\mathcal{T}/L^2),
   \end{equation}
   where \( f(z) \) is the function whose root corresponds to the value of \( z \).

3. \textbf{Numerical Solver:} Use \texttt{fsolve} to iteratively solve for \( z \) for a given value of \(t\). The solution requires:
   \begin{itemize}
       \item An initial guess for \( z \), which can be set based on physical intuition or previous computations.
       \item Appropriate tolerance settings to ensure convergence.
   \end{itemize}

4. \textbf{Iterative Computation:} For a range of \(t\), compute the corresponding values of \( z \). This produces the numerical solution \( z(t) \), which describes how \( z \) evolves with the rescaled time \(t\).

\subsection{ Numerical Behavior and Limits}

The behavior of \( z \) as a function of \(\mathcal{T}/L^2\) for the particle in a box example exhibits distinct regimes:

\begin{itemize}
    \item \textbf{For Large \(\mathcal{T}/L^2\):} As \(\mathcal{T}/L^2 \to \infty\), \(\mathcal{Z}(\cdot)\) decays exponentially (see Eq.\,(\ref{asymp1})):
    \begin{equation}
    \mathcal{Z}(\mathcal{T}/L^2) \sim 2 e^{-\frac{\pi^2}{4} \mathcal{T}/L^2}.
    \end{equation}
    In this regime, \( z \) converges to values consistent with the Gumbel statistics, reflecting the exponential tail of \(\mathcal{Z}(z)\).

    \item \textbf{For Small \(\mathcal{T}/L^2\):} As \(\mathcal{T}/L^2 \to 0\), \(\mathcal{Z}(\mathcal{T}/L^2)\) follows a power-law behavior (see Eq.\,(\ref{asymp1})):
    \begin{equation}
    \mathcal{Z}(\mathcal{T}/L^2) \sim (\pi \, \mathcal{T}/L^2)^{-\frac{1}{2}}.
    \end{equation}
    Here, \( z \) aligns with the Fréchet limit, capturing the influence of the heavy tail in the low-density regime.

    \item \textbf{Transition Around \(\mathcal{T}/L^2 \sim 1\):} Between these two asymptotic limits, we observe a smooth transition in \( z(\mathcal{T}/L^2) \). Around \(\mathcal{T}/L^2 \sim 1\), the behavior shifts from power-law to exponential decay, marking a crossover between the Fréchet and Gumbel regimes. This transition reflects the intermediate regime where neither tail fully dominates, and the full scaling function \(\mathcal{Z}(z)\) is required to describe the system accurately.
\end{itemize}

\begin{figure}[h]
    \centering
    \begin{minipage}[t]{0.49\linewidth}
        \centering
        \includegraphics[width=\linewidth]{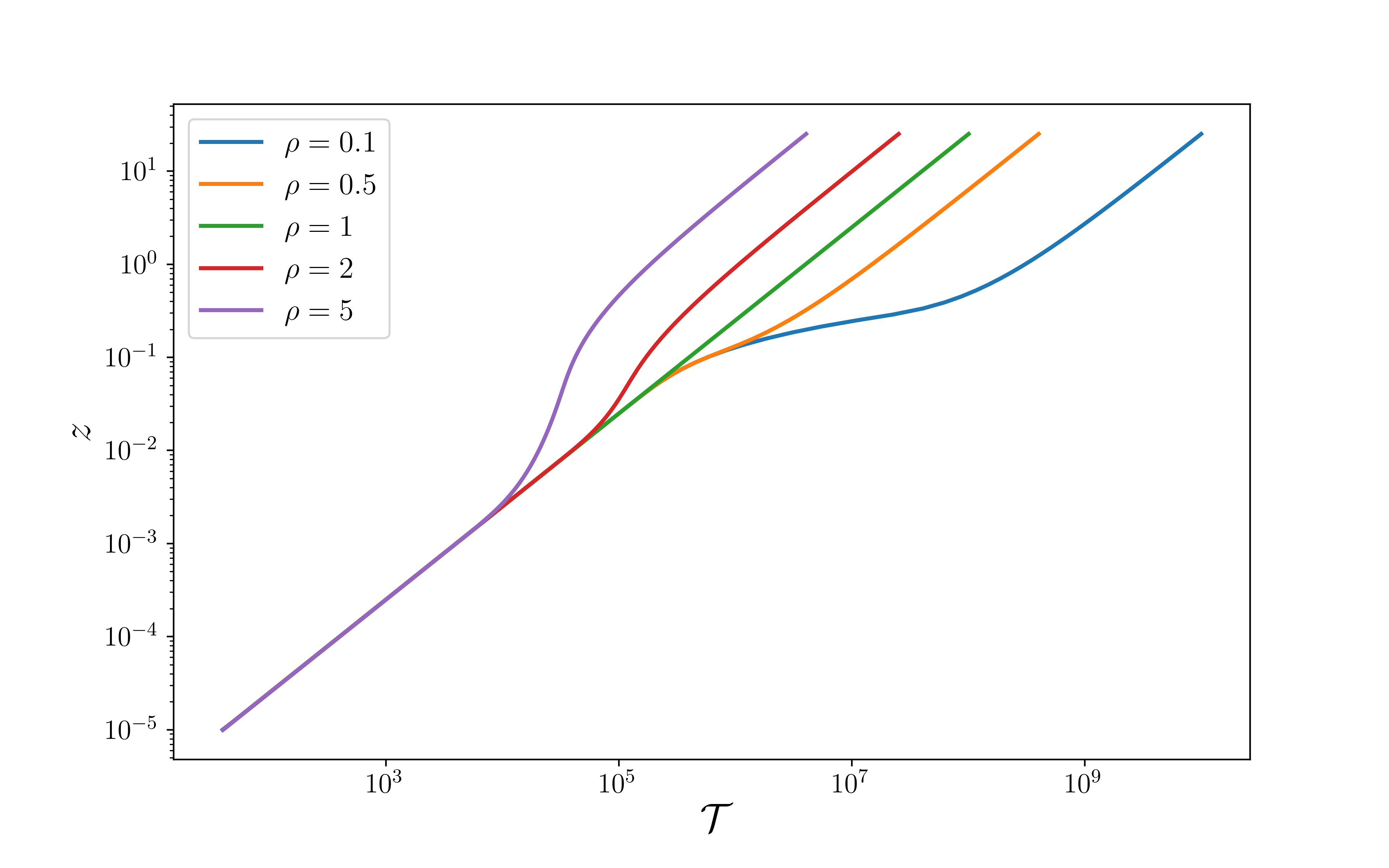}
    \end{minipage}
    \hfill
    \begin{minipage}[t]{0.49\linewidth}
        \centering
        \includegraphics[width=\linewidth]{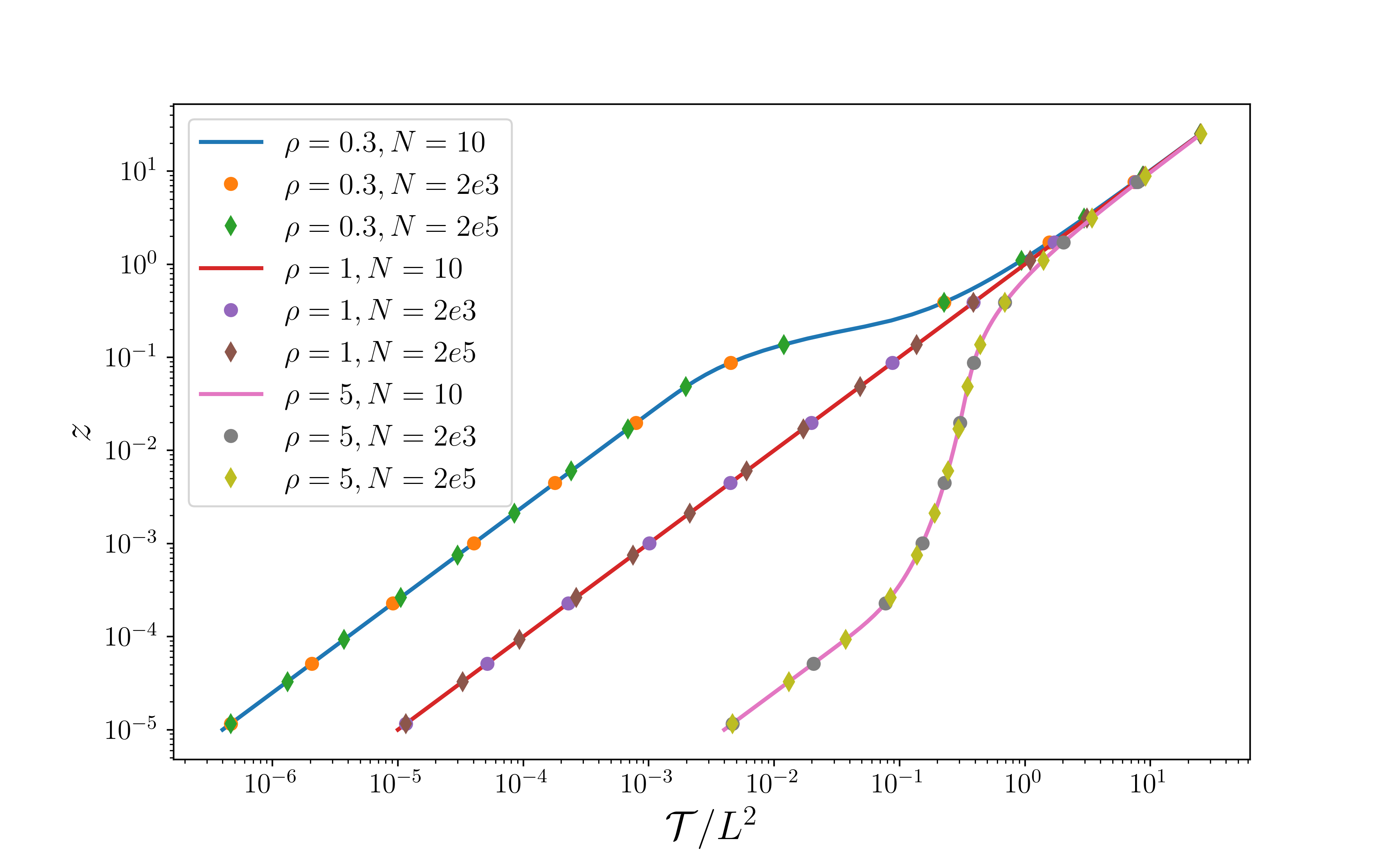}
    \end{minipage}
    \caption{The variable \( z \), defined by the solution of Eq.\,(\ref{eq:newzdef}). (Left) z as a function of \( \mathcal{T} \). Results for fixed particle number \( N = 2000 \) with varying \( \rho \). (Right) z as a function of \( \mathcal{T}/L^2 \) Results for varying \( N \) and $\rho$. Here we used $D=1/2$ and $x_0=1$.}\label{fig:zvstau}
\end{figure}

These limits and transitions are validated numerically by solving for \( z(t) \) and comparing the results to the expected asymptotic forms. The numerical results are presented in Fig.\,\ref{fig:zvstau}, showing \( z \) as a function of \(t\) for different densities \(\rho\). This highlights the consistency of our theory with the Gumbel and Fréchet distributions while demonstrating the intermediate behavior unique to our general solution.

\DIFaddbegin \section{\DIFadd{Effect of the Initial Position}}\label{appendix:initial}
\subsection{\DIFadd{Localized initial condition}}
\DIFadd{While our analysis in the main text focused on localized initial conditions with $x_0/L \rightarrow 0$, 
a more general case where the starting point scales as $x_0 = \alpha L$ with $0 < \alpha < 1$ can be treated within a similar framework. From Eq.\,(\ref{eq:exactp}), the FPT PDF of a single particle can be written as
}\begin{equation}\DIFadd{\label{appendeqp}
p(t|x_0) = \sum_{n=0}^\infty \frac{2 D \pi (n+\frac{1}{2})}{L^2} \sin\left(\pi (n+\frac{1}{2}) \alpha\right) e^{-D (\pi (n+\frac{1}{2}))^2 t / L^2},
}\end{equation}
\DIFadd{and therefore the survival probability is given by
}\begin{equation}\DIFadd{\label{appendeqs}
s(t|x_0) = \sum_{n=0}^\infty \frac{2}{\pi (n+\frac{1}{2})} \sin\left(\pi (n+\frac{1}{2}) \alpha\right) e^{-D (\pi (n+\frac{1}{2}))^2 t / L^2} \,.
}\end{equation}
\DIFadd{By retaining the full sine dependence in Eqs.\,(\ref{appendeqp}) and (\ref{appendeqs}) we capture how the survival probability depends on the relative position for any value of $\alpha$. When $\alpha \rightarrow 0$, expanding the sine term yields the limiting behavior discussed in the main text. A natural question arises: what happens when alpha is small but finite. 
}

\begin{figure}
    \centering
    \includegraphics[width=0.6\linewidth]{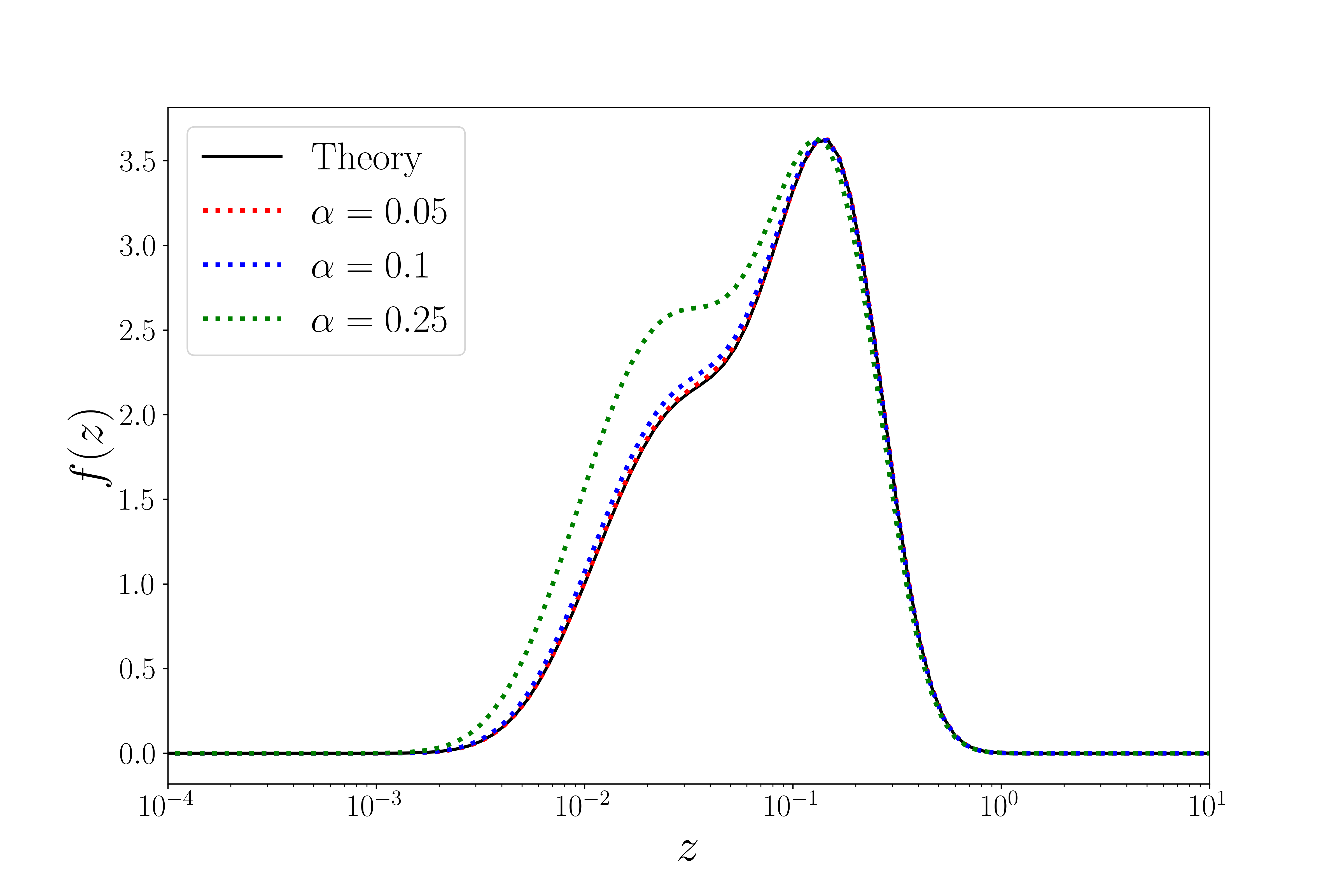}
    \caption{\DIFaddFL{PDF of the slowest FPT plotted as a function of the rescaled variable $z$ for different initial positions $x_0 = \alpha L$ with $\alpha = 0.05, 0.1, 0.25$ at fixed density $\rho = 1$ and $N = 2\times10^3$. 
Dots correspond to numerical results for the various values of $\alpha$, while the black solid line represents the theoretical prediction $f(z)$ from Eq.~(\ref{eq:theorypdf}). 
Deviations from the universal curve become noticeable only for $\alpha \sim 0.1$.
}}
    \label{fig:x0fig}
\end{figure}

\DIFadd{Figure~\ref{fig:x0fig} shows the probability density of the slowest first-passage time, plotted as a function of the rescaled variable $z$, 
for several values of $\alpha$ at fixed $\rho=1$ and $N=2\times10^3$. 
For small $\alpha$, the results collapse onto the universal theoretical curve $f(z)$ of Eq.~(\ref{eq:theorypdf}), 
whereas deviations become visible for $\alpha \sim 0.1$. In particular, the figure shows that the large-$z$ sector is less sensitive to $\alpha$ than the small-$z$ sector. The large-$z$ tail corresponds to extremely slow completion events, for which the detailed choice of initial position (encoded by $\alpha$) becomes less important.
}

\subsection{\DIFadd{Uniformly distributed initial condition}}

\DIFadd{Another option is to consider a uniform distribution of initial positions over the entire system, $x_0 = aL$ with $a \in [0,1]$. 
}

\DIFadd{Two distinct averaging procedures can be considered in this context. 
In the quenched case, the set of initial positions $\{x_i\}$ is fixed once and for all (for example, equally spaced points).
In contrast, in the annealed case, the initial positions are redrawn independently for each realization according to a uniform distribution, and the averaging is performed over both the diffusive trajectories and the random initial positions \mbox{
\cite{leibovich2013everlasting,ben2005transition}}\hskip0pt
.
The following subsections present explicit expressions for both cases.
}

\subsubsection{\DIFadd{Quenched Initial Conditions}}
\DIFadd{We consider the quenched setting where the $N$ initial positions are fixed deterministically at
}\begin{equation}
\DIFadd{x_i = \frac{iL}{N}, \qquad i = 1,\dots,N.
}\end{equation}
\DIFadd{For a single particle starting at $x_0$, the survival probability up to time $t$ is obtained using Eq.\,(\ref{eq:exactp}) and reads
}\begin{equation}
\DIFadd{s(\mathcal{T}|x_0) = \sum_{n=0}^{\infty} 
\frac{2}{\pi (n+\frac{1}{2})} 
\sin\!\left[\pi (n+\frac{1}{2})\frac{x_0}{L}\right]
\exp\!\left[-D(\pi(n+\frac{1}{2}))^2 \mathcal{T}/L^2\right].
}\end{equation}
\DIFadd{For this fixed configuration, the probability that all $N$ particles have reached the target by time $\mathcal{T}$ is
}\begin{equation}
\DIFadd{Q_N^{(\mathrm{q})}(\mathcal{T}) = 
\prod_{i=1}^{N}\!\left[1 - s\!\left(\mathcal{T}\,\Big|\,\frac{iL}{N}\right)\right].
\label{eq:Q_quenched_product}
}\end{equation}
\DIFadd{Taking the logarithm of Eq.\,(\ref{eq:Q_quenched_product}) gives
}\begin{equation}\DIFadd{\label{eq:approxapendx}
\ln Q_N^{(\mathrm{q})}(\mathcal{T})
= \sum_{i=1}^{N} \ln\!\left[1 - s\!\left(\mathcal{T}\,\Big|\,\frac{iL}{N}\right)\right].
}\end{equation}
\DIFadd{In the long-time limit, where $s(\mathcal{T}|x_0)\ll 1$, we expand 
\begin{equation}\DIFadd{
\sum_{i=1}^{N} \ln\!\left[1 - s\!\left(\mathcal{T}\,\Big|\,\frac{iL}{N}\right)\right]
\simeq - \sum_{i=1}^{N} s\!\left(\mathcal{T}\,\Big|\,\frac{iL}{N}\right).
}\end{equation}
Then, in the large-$N$ limit, the discrete sum can be replaced by an integral, gives
}\begin{equation}
\DIFadd{Q_N^{(\mathrm{q})}(\mathcal{T})
\simeq \exp\!\left[-N \int_0^1 s(\mathcal{T}|\alpha L )\,d\alpha\right].
\label{eq:Q_quenched_continuum}
}\end{equation}
\DIFadd{Substituting the explicit form of $s(t|x_0)$ and performing the spatial integration yields
}\begin{equation}
\DIFadd{Q_N^{(\mathrm{q})}(\mathcal{T})
\simeq 
\exp\!\left\{
-N
\sum_{n=0}^{\infty}
\frac{2}{\pi^2 (n+\frac{1}{2})^2}
\exp\!\left[-D(\pi(n+\frac{1}{2}))^2 \mathcal{T}/L^2\right]
\right\}.
\label{eq:Q_quenched_final}
}\end{equation}
\DIFadd{In the last step we have used 
$\int_0^1 \sin[\pi(n+1/2)a]\, da = 1/[\pi(n+1/2)]$ for integer $n$. }

\DIFadd{For $\mathcal{T} \to \infty$ (i.e., $\mathcal{T} \gg L^2/D$), $\bar{s}(\mathcal{T})$ is dominated by the slowest decaying mode $n=0$,
}\begin{equation}
\DIFadd{\bar{s}(\mathcal{T})
\simeq
\frac{8}{\pi^2}
\exp\!\left( - \frac{\pi^2}{4} \mathcal{T}\,D/L^2 \right),
\qquad
\mathcal{T} \to \infty.
\label{eq:S_long}
}\end{equation}
\DIFadd{Hence, 
}\begin{equation}
\DIFadd{Q_N(\mathcal{T})
\simeq
\exp\!\left[
- N \,\frac{8}{\pi^2} 
\exp\!\left( - \frac{\pi^2}{4} \mathcal{T}\,D/L^2 \right)
\right],
\qquad
\mathcal{T} \to \infty.
\label{eq:P_long}
}\end{equation}
\DIFadd{Note that here we assumed $\mathcal{T}\,D/L^2 \gg 1$ and $N\gg 1$ such that $N\,\frac{8}{\pi^2} 
\exp\!\left( - \frac{\pi^2}{4} \mathcal{T}\,D/L^2 \right)\propto    O(1)$. Eq.\,(\ref{eq:P_long}) is a exponential form of the type 
$\exp[-A \exp(-B\tau)]$, 
which is the standard Gumbel-like tail. 
}

\subsubsection{\DIFadd{Annealed initial condition }}
\begin{figure}
    \centering
    \includegraphics[width=0.6\linewidth]{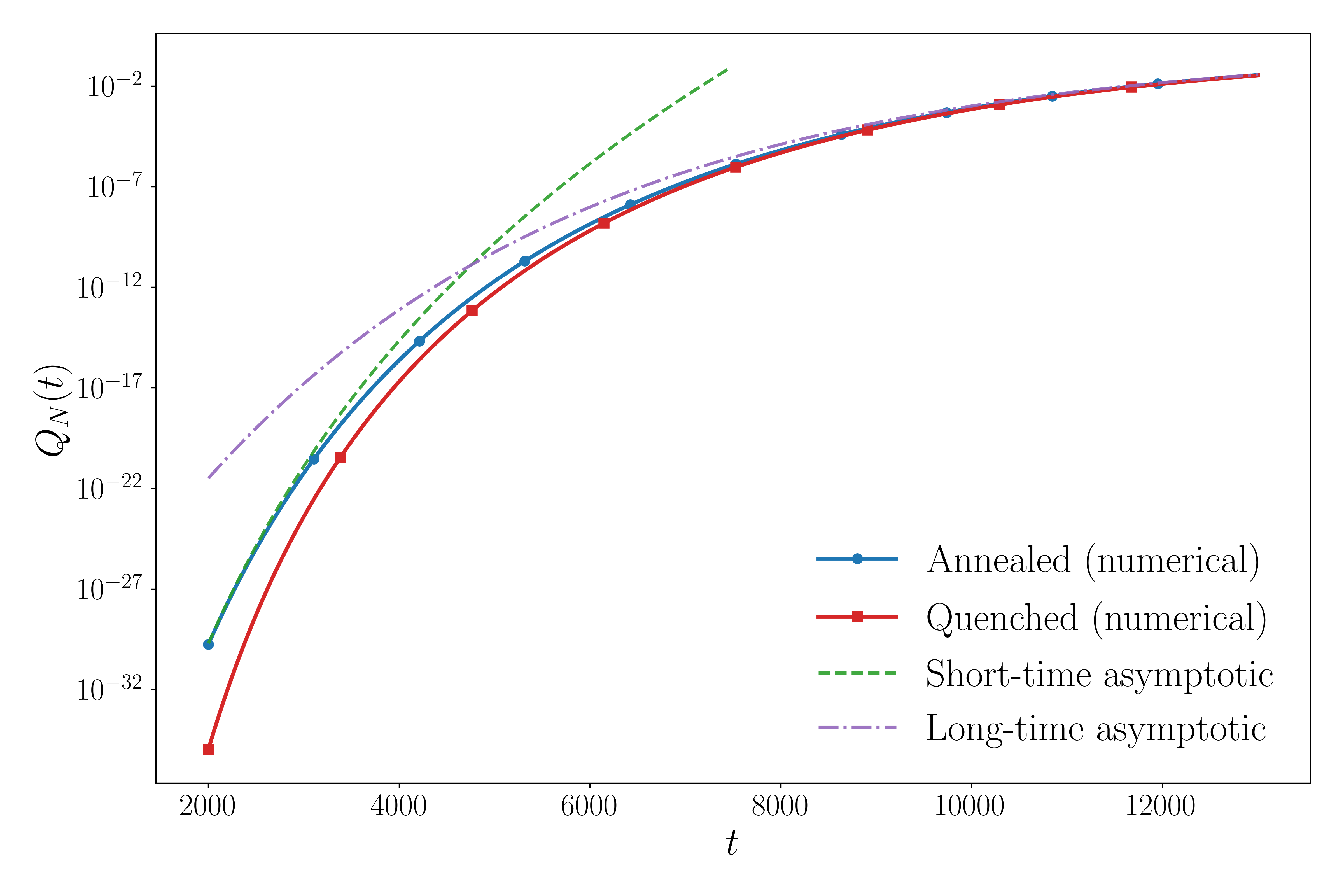}
    \caption{\DIFaddFL{Comparison between the quenched and annealed formulations of the slowest first-passage probability 
        $Q_N(t)$ for a uniform distribution of initial positions. 
        The quenched curve is computed using the exact product in Eq.~\eqref{eq:Q_quenched_product}, while the annealed curve follows 
    Eq.~\eqref{eq:Q_annealed1}. 
    The dashed green and dotted purple lines represent the short- and long-time asymptotic limits, respectively, given in Eqs.\,(\ref{eq:appendixshortlim}) and (\ref{eq:P_long}). Here we used $N=L=100$.
}}
    \label{fig:annealedvsquenched}
\end{figure}

\DIFadd{In the annealed setting, each particle begins at a random position 
$x_0 = aL$ with $a \in [0,1]$, drawn independently and uniformly for every realization.  
The survival probability of a single particle is then averaged over these possible 
initial positions,
}\begin{equation}\DIFadd{\label{eq:appends}
\overline{s}(\mathcal{T}) = \int_0^1 d\alpha\, s(\mathcal{T}|\alpha L)
= \sum_{n=0}^{\infty}
\frac{2}{\pi^2 (n+\frac{1}{2})^2}
\exp\!\left[-D(\pi(n+\frac{1}{2}))^2 \mathcal{T}/L^2\right].
}\end{equation}
\DIFadd{Since all particles are independently sampled from the same distribution, the probability 
that all have reached the boundary by time $t$ reads
}\begin{equation}
\DIFadd{Q_N^{(\mathrm{a})}(\mathcal{T})
= \Big[\,1 - \overline{s}(\mathcal{T})\,\Big]^N.
\label{eq:Q_annealed1}
}\end{equation}
\DIFadd{which for large $N$ and $\bar{s}(\mathcal{T})\propto O(1/N)$, gives
}\begin{equation}
\DIFadd{Q_N^{(\mathrm{a})}(\mathcal{T})
\simeq
\exp\!\left[-N\,\overline{s}(\mathcal{T})\right].
\label{eq:Q_annealed}
}\end{equation}
\DIFadd{Eq.~\eqref{eq:Q_annealed} becomes
}\begin{equation}
\DIFadd{Q_N^{(\mathrm{a})}(\mathcal{T})
\simeq
\exp\!\left[-N 
\sum_{n=0}^{\infty}
\frac{2}{\pi^2 (n+\frac{1}{2})^2}
\exp\!\left[-D(\pi(n+\frac{1}{2}))^2 \mathcal{T}/L^2\right]\right].
}\end{equation}
\DIFadd{This expression is identical to the quenched result, therefore we expect converges of the two solutions for long times. Performing the same asymptotic expansion for $\mathcal{T}\rightarrow \infty$ as in the quenched case yields the expression in Eq.~(C8). 
Therefore, in the limit of large $T$, the annealed and quenched solutions converge and become indistinguishable, as shown in Fig.\,\ref{fig:annealedvsquenched}. Consequently, the asymptotic limit $T\to\infty$ corresponds to the same Gumbel form for both ensembles. A comparison between the two cases for finite N and L is left for future work.
}

\DIFadd{To analyze on the other hand the short-time behavior, we note that the approximation used in Eq.~(\ref{eq:Q_annealed}),
\((1-s)^N \simeq \exp(-N s)\), is no longer valid. 
This expansion assumes \(s \ll 1\), whereas at short times the single-particle survival probability \(s(t)\) is close to unity. }
\DIFadd{For $\mathcal{T}\!\ll\! L^2/D$, define $a = D\pi^2\mathcal{T}/L^2$ and approximate the sum in Eq.\,(\ref{eq:appends}) 
}\begin{align}
\DIFadd{F(a)
}&\DIFadd{=\sum_{n=0}^\infty 
\frac{2}{\pi^2 (n+\tfrac{1}{2})^2}
e^{-a (n+\tfrac{1}{2})^2}
\simeq
1-
\frac{2}{\pi^2}
\!\int_0^\infty\!
\frac{1-e^{-a x^2}}{x^2}\,dx
\simeq
1-\frac{2}{\pi^{3/2}}\sqrt{a},
\label{eq:F_shorttime}
}\end{align}
\DIFadd{which yields
}\begin{equation}
\DIFadd{\bar{s}(\mathcal{T})
\simeq
1-\frac{2}{\sqrt{\pi}}\frac{\sqrt{D\mathcal{T}}}{L},
\qquad
\mathcal{T}\ll\frac{L^2}{D}.
}\end{equation}
\DIFadd{Inserting this into  $Q_N^{(\mathrm{a})}(\mathcal{T})=(1-\bar{s}(\mathcal{T})^N$
gives the short-time asymptotic
}\begin{equation}\label{eq:appendixshortlim}
  \DIFadd{Q_N^{(a)}(\mathcal{T}) \simeq \left(2\sqrt{\frac{D\,\mathcal{T}}{\pi L^2}} \right)^N ,\qquad\mathcal{T} \to 0.
}\end{equation}

\DIFadd{As discussed above, both the quenched and annealed formulations ultimately converge to the same expression in the long-time limit. 
The comparison in Fig.~\ref{fig:annealedvsquenched} illustrates how the various asymptotic expressions behave across the entire temporal range.
For large times, Eq.~(\ref{eq:P_long}) coincides with both the quenched and annealed results, indicating that all three curves collapse onto a common asymptotic form. At short times, the annealed curve overlap with the short-time approximation from Eq.~(\ref{eq:appendixshortlim}), whereas the quenched result shows a visible deviation. 
However, in this regime the probability itself becomes exceedingly small, reaching values of order $O(10^{-32})$ for the parameters used here. 
Such exponentially small probabilities correspond to extremely unlikely events—the cases where the slowest particle among many happens to reach the target at an anomalously short time. 
This vanishingly small weight renders the short-time discrepancy physically irrelevant, since these events are practically unobservable. 
This small numerical values are in fact consistent with intuition: when the initial positions are uniformly distributed, some particles start far from the target, and hence the probability that all particles have already arrived by a very short time is very small. 
Consequently, for the uniform initial condition, unlike the example discussed in the main text, this case involves only one physically relevant limiting function, as it is dominated by the Gumbel distribution. }
\DIFadd{For a broader discussion of quenched versus annealed averaging in diffusion and first-passage problems,  
see ~\mbox{
\cite{leibovich2013everlasting,ben2005transition,Donsker1975,Donsker1979,Hartmann2023,Burenev2024,Madrid2020}}\hskip0pt
.
}

\section{\DIFadd{Derivation of the Extreme First-Passage time moments}}\label{app:mefpt}
\subsection{\DIFadd{Derivation of the Mean Extreme First-Passage Time}}
\DIFadd{Starting from Eq.~(\ref{eq:defmean1}) and using the definition of the Jacobi elliptic theta function $ \vartheta_2(e^{-\pi^2 t})=2\sum_{n=0}^\infty e^{-\pi^2(n+1/2)^2 t}$, we can rewrite Eq.\,(\ref{eq:defmean1}) as
}\begin{align}
\DIFadd{\langle \mathcal{T} \rangle }& \DIFadd{\approx\frac{L^2}{D} 2x_0 \rho \sum_{n=0}^\infty \pi^2(n+1/2)^2\int_0^\infty \tau\, e^{-\pi^2(n+1/2)^2 \tau} \prod_{m=0}^\infty\exp{\left[-2x_0 \rho\, e^{-\pi^2(m+1/2)^2\,\tau}\right]} d\tau }\\&\DIFadd{=\frac{L^2}{D}2 x_0 \rho \sum_{n=0}^\infty \frac{1}{\pi^2(n+1/2)^2} \int_0^\infty y\, e^{-y} \prod_{m=0}^\infty\exp{\left[-2x_0 \rho\, e^{-\frac{(m+1/2)^2}{(n+1/2)^2}\,y}\right]} dy \nonumber}\\
&\DIFadd{=\frac{L^2}{D}2 x_0 \rho \sum_{n=0}^\infty \frac{1}{\pi^2(n+1/2)^2} \int_0^\infty y\, e^{-y-2x_0\rho e^{-y}} \prod_{m\neq n}^\infty\exp{\left[-2x_0 \rho\, e^{-\frac{(m+1/2)^2}{(n+1/2)^2}\,y}\right]} dy \nonumber}\\
&\DIFadd{=\frac{L^2}{D}2 x_0 \rho \sum_{n=0}^\infty \frac{1}{\pi^2(n+1/2)^2} \int_0^\infty y\, e^{-y-2x_0\rho e^{-y}}  R_n(y)dy \nonumber
}\end{align}
\DIFadd{which is Eq.\,(\ref{eq:mfpt1}) in the main text.
}

\subsection{\DIFadd{Extreme FPT variance }}
\DIFadd{The second moment of the extreme FPT is defined by
}\begin{equation}\DIFadd{\label{eq:secondmom}
    \langle \mathcal{T}^2\rangle = \int_0^{\infty} q_N(\mathcal{T}') \, \mathcal{T}'^2 \, d\mathcal{T}',
}\end{equation}
\DIFadd{then in the small-density limit, using Eqs.\,(\ref{eq:momqnremind}) and (\ref{eq:mfpt1}) we can write Eq.\,(\ref{eq:secondmom}) as
}\begin{align}
\DIFadd{\langle \mathcal{T}^2 \rangle
}&\DIFadd{\approx\left(\frac{L^2}{D}\right)^22 x_0 \rho \sum_{n=0}^\infty \frac{1}{\pi^4(n+1/2)^4} \int_0^\infty y^2\, e^{-y-2x_0\rho e^{-y}}  R_n(y)dy}\\& \DIFadd{\approx \left(\frac{L^2}{D}\right)^22 x_0 \rho \sum_{n=0}^\infty \frac{1}{\pi^4(n+1/2)^4} \int_0^\infty y^2\, e^{-y-x_0\rho \sqrt{\pi/y}}  dy \,.
}\end{align}
\DIFadd{Using the identity $\sum_{n=0}^\infty \frac{1}{\pi^4(n+1/2)^4}=1/6$ and performing the integration over $y$, we find
}\begin{equation}\DIFadd{\label{eq:lowdenmfpt1}
     \langle \mathcal{T}^2 \rangle\approx \left(\frac{L^2}{D}\right)^2 x_0 \rho \frac{2}{6\sqrt{\pi}} \, G^{0,0}_{3,0} \left( \frac{(x_0\rho) ^2 \pi}{4} \,\middle|\, \begin{matrix} 0,\; \tfrac{1}{2},\; 3 \\ \end{matrix} \right)\, .
}\end{equation}
\DIFadd{where $G(\cdot)$ is the Meijer G-function. Using the small-argument asymptotics of the Meijer G-function %
\mbox{
$G^{0,0}_{3,0} \left( \frac{x^2 \pi}{4} \,\middle|\, \begin{matrix} 0,\; \tfrac{1}{2},\; 3 \\ \end{matrix} \right)\sim 2\sqrt{\pi}$
}
, Eq.\,(\ref{eq:lowdenmfpt1}) simplify to
}\begin{equation}\DIFadd{\label{eq:mefptlowrho2}
    \langle \mathcal{T}^2 \rangle \rightarrow \frac{2}{3}\frac{\rho\,L^4 x_0}{D^2}\, \, .
    }\end{equation}

\DIFadd{The variance is defined as
}\begin{equation}
    \DIFadd{Var(\mathcal{T})\equiv  \langle \mathcal{T}^2 \rangle- \langle \mathcal{T} \rangle^2\, .
}\end{equation}
\DIFadd{Using Eqs.\,(\ref{eq:mefptlowrho}) and (\ref{eq:mefptlowrho2}) we find
}\begin{equation}\DIFadd{\label{eq:varloeden}
Var(\mathcal{T})\approx\left(\frac{\,L^2}{D}\right)^2\left(\frac{2}{3}\rho \,x_0 - \rho^2x_0^2\right) \, .
}\end{equation}

\begin{figure}
    \centering
    \includegraphics[width=0.6\linewidth]{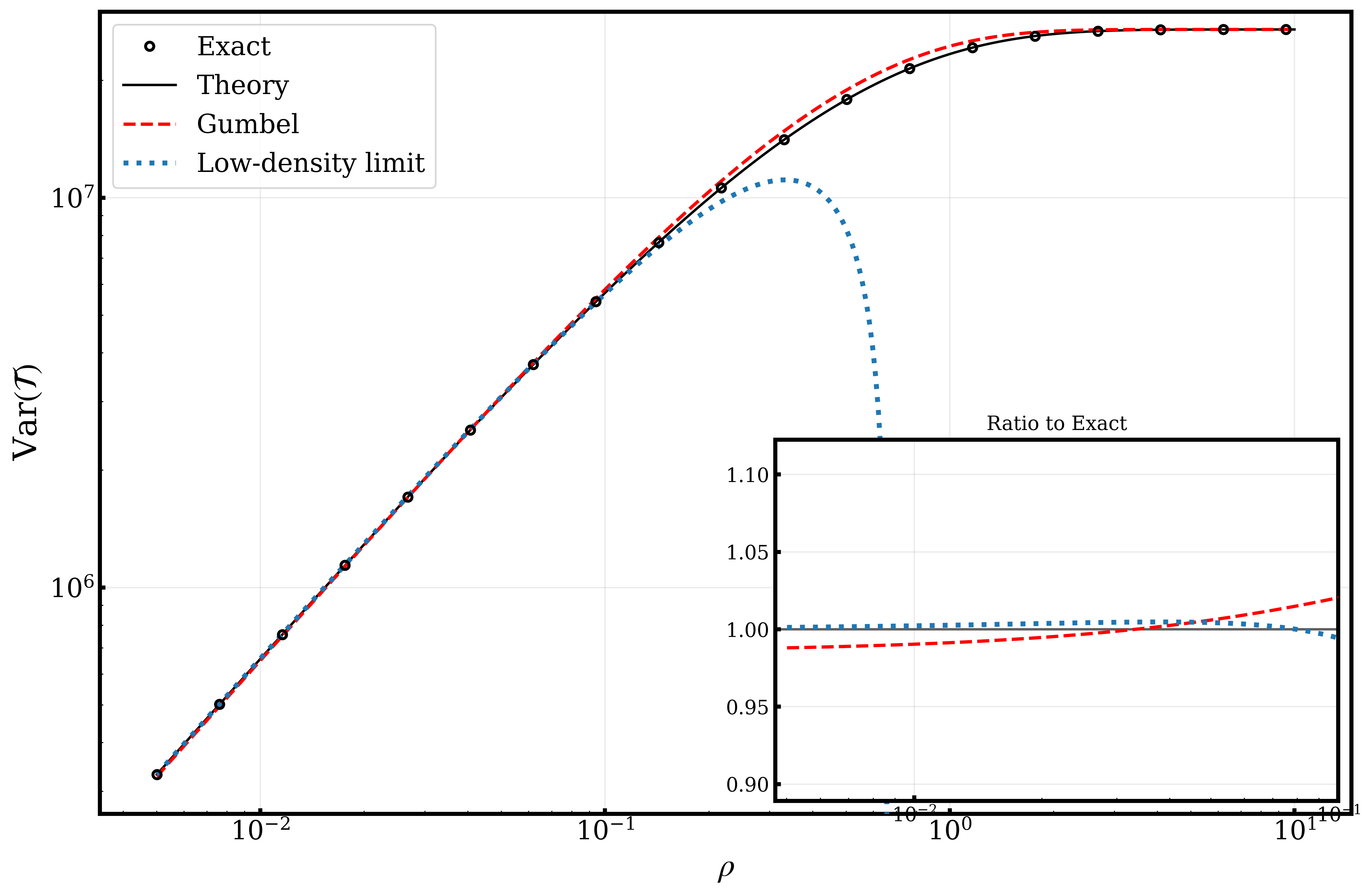}
    \caption{\DIFaddFL{Variance of the extreme FPT versus density $\rho$. Black line: theory; open circles: exact numeric integration; red dashed: Gumbel approximation (high density limit) using Eqs.\,(\ref{eq:meangumb}) and (\ref{eq:gumbsecmom}); blue dotted: low-density limit in Eq.\,(\ref{eq:varloeden}).
Inset: ratio of approximate to exact variance, showing convergence in the low-$\rho$ regime and small deviations near $\rho \rightarrow 0$. Here we used fixed $L=100$ and changed N accordingly.}}
    \label{fig:var_compare}
\end{figure}

\DIFadd{In the high-density limit, using the result in Eq.\,(\ref{eq:momgumb}) for $n=2$, we can write the second moment as
}\begin{align}\DIFadd{\label{eq:gumbsecmom}
\langle \mathcal{T}^2 \rangle
}&\DIFadd{\approx2 x_0 \rho \left(\frac{4}{\pi^2}\right)^2 \left(\frac{L^2}{D}\right)^2 \int_0^\infty y^2 \exp\left[-y - 2\,x_0 \rho\, e^{-y}\right] dy}\\& \DIFadd{=  4\, x_0 \rho \left(\frac{4}{\pi^2}\right)^2 \left(\frac{L^2}{D}\right)^2 \,_2F_2\big(\{1,1,1\}; \{2,2,2\}; -2x_0 \rho \big) \,.
}\end{align}
\DIFadd{namely the second moment converges to the one calculated from the Gumbel's distribution.
Combining with Eq.~(\ref{eq:meangumb}) in the main text, we find that in the large–density limit the variance approaches its asymptotic Gumbel value, 
while for small~$\rho$ it exhibits a slight deviation captured by the low–density expansion. Fig.~\ref{fig:var_compare} shows that even at small~$\rho$ the analytic low–density approximation closely follows the exact numerical result, 
with only a small deviation compared to the Gumbel limit.
}

\DIFaddend \section{Supplementary Technical Details for Diffusion on Fractal Geometry }\label{app:othermodels}

Consider a compact random walk on a fractal domain of
fractal dimension $d_f$ and walk dimension
$d_w$, following the O’Shaughnessy–Procaccia transport equation \cite{o1985analytical,o1985diffusion}. The solution is defined in the domain $r\in[0,L]$ with an absorbing inner boundary at the target, $P(r=0,t)=0$, and a reflecting outer boundary at $r=R$, i.e. zero radial flux $r^{\,d_f-d_w+1}\partial_r P|_{r=R}=0$. The survival probability exhibits the scaling form in Eq.\,(\ref{eq:gens}) \cite{baravi2025first}
\begin{equation}\label{eq:fractal_scaling}
   S(t)\sim
   \Bigl(\frac{r_0}{L}\Bigr)^{(d_w-d_f)}
   \mathcal Z\!\Bigl(\frac{t}{\tau_L}\Bigr),
   \qquad
   \tau_L = L^{d_w}/K ,
\end{equation}
where $K$ is the generalized diffusion coefficient defined in Eq.\,(\ref{eq:diff_frac}).
The persistence exponent is $\theta = 1 - d_f/d_w$ and $d_w\theta = d_w-d_f$ \cite{havlin2002diffusion}.

\par The infinite–density $\,\mathcal I(\tau)$ and its integral
$\mathcal Z(\tau)$, obtained from the eigenmode spectrum of
the O’Shaughnessy–Procaccia operator, are \cite{meyer2011universality,baravi2025solutions}
\begin{align} 
  \mathcal I(\tau) &=
     \frac{2^{2\nu-3}\,d_w}{\Gamma(2-\nu)}
     \sum_{n=0}^{\infty}
        \frac{J_\nu(z_{-\nu,n})}{J_{1-\nu}(z_{-\nu,n})}\,
        z_{-\nu,n}^{\,3-2\nu}\,
        e^{-\,d_w^{2}z_{-\nu,n}^{2}\tau/4},
     \label{eq:I_fractal2}\\[4pt]
  \mathcal Z(\tau) &= \int_{\tau}^{\infty}\!\mathcal I(u)\,du,
     \qquad  \nu \equiv \frac{d_f}{d_w}<1,
     \label{eq:Z_fractal}
\end{align}
where $z_{-\nu,n}$ are the positive zeros of
$J_{-\nu}(z)$, with $\nu = d_f/d_w$. Note that Eq.~(\ref{eq:I_fractal2}) is an asymptotic (large-$\tau$) scaling form in a finite domain and is not normalized. A complete, uniformly valid description requires, in addition, the infinite-domain scaling solution; matching the two asymptotics yields a uniform solution of the problem \cite{baravi2025solutions}.

\subsection*{Short– and long–time limits}

Using $J_\alpha(z)\simeq(z/2)^\alpha/\Gamma(1+\alpha)$ for
$z\!\to\!0$ and replacing the sum by an integral for
$\tau\!\ll\!1$, one finds
\begin{equation}
   \mathcal I(\tau)\;\xrightarrow{\tau\ll1}\;
   \frac{d_w^{\,\nu-2}}{2\,\Gamma(\nu)}\;\tau^{-\,\!(2-\nu)},
   \qquad
   \mathcal Z(\tau)\;\xrightarrow{\tau\ll1}\;
   \frac{\tau^{-\,\!(1-\nu)}}{\Gamma(2-\nu)} .
\end{equation}
Thus $\mathcal Z(\tau)\propto\tau^{-\theta}$ with
$\theta=1-\nu=1-d_f/d_w$, in agreement with
Eq.\,\eqref{eq:fractal_scaling}.  

For $\tau\!\gg\!1$ the lowest eigenvalue $z_{-\nu,0}$ dominates,
giving the exponential cut–off
\begin{equation}\label{eqappend:c5}
   \mathcal Z(\tau)\;\xrightarrow{\tau\gg1}\;
   C\,\exp\!\bigl[ -\,\lambda_0 \tau \bigr],
   \quad
   \lambda_0 = \frac{d_w^{2}z^{2}_{-\nu,0}}{4},\;
   C=\frac{2^{2\nu-3}d_w J_\nu(z_{-\nu,0})}{\lambda_0\,\Gamma(2-\nu)J_{1-\nu}(z_{-\nu,0})}.
\end{equation}
As emphasized in the main text, the finite-size–induced exponential cutoff of the single-particle FPT (Eq.(\ref{eqappend:c5})) implies that, in the high-density limit ( $L$, $N\to\infty$ at fixed $\rho=N/L^{d_f}$), the extreme (slowest) FPT converges to the Gumbel law (consistent with Eq.(\ref{eq:gumbelpdf})).

\section{Simulation on the Sierpinski gasket}\label{appendix:simulation}
The confining length scale $L$ is naturally defined via the total number of sites $n$ in the system. On the Sierpiński gasket, the number of nodes at iteration (generation) $m$ grows as $n = (3 + 3^{m+1})/2$ \cite{teguia2005sierpi}. The linear length $L$ is then given by $L = n^{1/d_f}$

The master equation for the probability $P_n(i)$ for a random walker on a graph $G$ to be at site $i$ at time $n$ is given in general by
   \begin{equation}
       P_{n+1}(i)=A\,P_{n}(i)
    \end{equation}
    where $A$ is the adjacency matrix. Apply an absorbing boundary condition at the target site $i=i_{target}$, namely
     \begin{equation}
       P_{n}(i_{target})=0.
    \end{equation}
    The survival probability, namely the probability that the particle did not arrive to the target by the $n$-th jump is given by
    \begin{equation}\label{eq:survsirp}
        S_n=\sum_{i} P_n(i),
    \end{equation}
    then the FPT probability distribution is defined by the numeric derivative of Eq. (\ref{eq:survsirp}) as
 \begin{equation}
        \eta_n=\frac{S_{n}-S_{n-1} }{\Delta n}, 
    \end{equation}
    where $\Delta n=1$.

\DIFaddbegin


\begin{thebibliography}{10}


\bibitem{schuss2019redundancy}
Z.~Schuss, K.~Basnayake, and D.~Holcman.
\newblock Redundancy principle and the role of extreme statistics in molecular
  and cellular biology.
\newblock {\em Physics of life reviews}, 28:52--79, 2019.

\DIFaddbegin \bibitem{majumdar2024statistics}
\DIFadd{S.~N. Majumdar and G.~Schehr.
}\newblock {\em \DIFadd{Statistics of Extremes and Records in Random Sequences}}\DIFadd{.
}\newblock \DIFadd{Oxford University Press, 2024.
}

\DIFaddend \bibitem{lawley2023slowest}
S.~D. Lawley and J.~Johnson.
\newblock Slowest first passage times, redundancy, and menopause timing.
\newblock {\em Journal of Mathematical Biology}, 86(6):1--53, 2023.

\bibitem{guerin2016mean}
T.~Gu{\'e}rin, N.~Levernier, O.~B{\'e}nichou, and R.~Voituriez.
\newblock Mean first-passage times of non-markovian random walkers in
  confinement.
\newblock {\em Nature}, 534(7607):356--359, 2016.

\bibitem{jenkinson1955frequency}
A.~F. Jenkinson.
\newblock The frequency distribution of the annual maximum (or minimum) values
  of meteorological elements.
\newblock {\em Quarterly Journal of the Royal meteorological society},
  81(348):158--171, 1955.

\bibitem{novak2011extreme}
S.~Y. Novak.
\newblock Extreme value methods with applications to finance.
\newblock {\em Monographs on Statistics and Applied Probability}, 122:22, 2011.


\bibitem{majumdar2002extreme}
S.~N. Majumdar and P.~L. Krapivsky.
\newblock Extreme value statistics and traveling fronts: Application to
  computer science.
\newblock {\em Physical Review E}, 65(3):036127, 2002.

\bibitem{fortin2015applications}
J.-Y. Fortin and M.~Clusel.
\newblock Applications of extreme value statistics in physics.
\newblock {\em Journal of Physics A: Mathematical and Theoretical},
  48(18):183001, 2015.

\bibitem{biroli2025stronglycorrelatedstochasticsystems}
M.~Biroli.
\newblock Strongly correlated stochastic systems.
\newblock {\em arXiv preprint arXiv:2508.12818}, 2025.

\bibitem{matsinos2024extreme}
M.~Evangelos.
\newblock Extreme-value statistics: Rudiments and applications.
\newblock {\em arXiv preprint arXiv:2407.00725}, 2024.

\bibitem{holl2020extreme}
M.~H{\"o}ll, W.~Wang, and E.~Barkai.
\newblock Extreme value theory for constrained physical systems.
\newblock {\em Physical Review E}, 102(4):042141, 2020.

\bibitem{albeverio2006extreme}
S.~Albeverio, V.~Jentsch, and H.~Kantz.
\newblock {\em Extreme events in nature and society}.
\newblock Springer Science \& Business Media, 2006.

\bibitem{embrechts2013modelling}
P.~Embrechts, C.~Kl{\"u}ppelberg, and T.~Mikosch.
\newblock {\em Modelling extremal events: for insurance and finance},
  volume~33.
\newblock Springer Science \& Business Media, 2013.

\bibitem{huang2025first}
F.~Huang and H.~Chen.
\newblock First-passage and extreme value statistics for overdamped brownian
  motion in a linear potential.
\newblock {\em Physica A: Statistical Mechanics and its Applications}, page
  130673, 2025.

\bibitem{bouchaud1997universality}
J.~P. Bouchaud and M.~M{\'e}zard.
\newblock Universality classes for extreme-value statistics.
\newblock {\em Journal of Physics A: Mathematical and General}, 30(23):7997,
  1997.

\bibitem{flandoli2025extreme}
F.~Flandoli, S.~Galatolo, P.~Giulietti, and S.~Vaienti.
\newblock Extreme value theory and poisson statistics for discrete time
  samplings of stochastic differential equations.
\newblock {\em Communications in Mathematical Physics}, 406(10):231, 2025.

\bibitem{gumbel1958statistics}
E.~J. Gumbel.
\newblock {\em Statistics of extremes}.
\newblock Columbia university press, 1958.

\bibitem{fisher1928limiting}
R.~A. Fisher and L.~H.~C. Tippett.
\newblock Limiting forms of the frequency distribution of the largest or
  smallest member of a sample.
\newblock In {\em Mathematical proceedings of the Cambridge philosophical
  society}, volume~24, pages 180--190. Cambridge University Press, 1928.

\bibitem{gnedenko1943distribution}
B.~Gnedenko.
\newblock Sur la distribution limite du terme maximum d'une serie aleatoire.
\newblock {\em Annals of mathematics}, 44(3):423--453, 1943.

\bibitem{majumdar2020extreme}
S.~N. Majumdar, A.~Pal, and G.~Schehr.
\newblock Extreme value statistics of correlated random variables: a
  pedagogical review.
\newblock {\em Physics Reports}, 840:1--32, 2020.

\bibitem{hall1979rate}
P.~Hall.
\newblock On the rate of convergence of normal extremes.
\newblock {\em Journal of Applied Probability}, 16(2):433--439, 1979.

\bibitem{gyorgyi2008finite}
G.~Gy{\"o}rgyi, N.~R. Moloney, K.~Ozog{\'a}ny, and Z.~R{\'a}cz.
\newblock Finite-size scaling in extreme statistics.
\newblock {\em Physical review letters}, 100(21):210601, 2008.

\bibitem{majumdar2024decorrelation}
S.~N. Majumdar and G.~Schehr.
\newblock Decorrelation of a leader by an increasing number of followers.
\newblock {\em Physical Review E}, 110(4):044111, 2024.

\bibitem{zarfaty2022discrete}
L.~Zarfaty, E.~Barkai, and D.~A. Kessler.
\newblock Discrete sampling of extreme events modifies their statistics.
\newblock {\em Physical Review Letters}, 129(9):094101, 2022.

\bibitem{zarfaty2021accurately}
L.~Zarfaty, E.~Barkai, and D.~A. Kessler.
\newblock Accurately approximating extreme value statistics.
\newblock {\em Journal of Physics A: Mathematical and Theoretical},
  54(31):315205, 2021.

\bibitem{mikosch2020gumbel}
T.~Mikosch and J.~Yslas.
\newblock Gumbel and {F}r{\'e}chet convergence of the maxima of independent
  random walks.
\newblock {\em Advances in Applied Probability}, 52(1):213--236, 2020.

\bibitem{oshanin2013anomalous}
G.~Oshanin, A.~Rosso, and Gr{\'e}gory Schehr.
\newblock Anomalous fluctuations of currents in sinai-type random chains with
  strongly correlated disorder.
\newblock {\em Physical Review Letters}, 110(10):100602, 2013.

\bibitem{fn:GEVcomment}
The GEV CDF with location $\mu$, scale $\sigma>0$, and shape $\kappa$ is
  $G_\kappa(z)=\exp\!\{-[1+\kappa(z-\mu)/\sigma]^{-1/\kappa}\}$, which
  continuously recovers the Gumbel ($\kappa\to 0$), Fr\'echet ($\kappa>0$), and
  Weibull ($\kappa<0$) limits.

\bibitem{de2018superstatistical}
C.~De Michele and F.~Avanzi.
\newblock Superstatistical distribution of daily precipitation extremes: A
  worldwide assessment.
\newblock {\em Scientific reports}, 8(1):14204, 2018.

\bibitem{kotz2000extreme}
K.~Samuel and N.~Saralees.
\newblock {\em Extreme value distributions: theory and applications}.
\newblock World Scientific, 2000.

\bibitem{singh1998generalized}
V.~P. Singh.
\newblock Generalized extreme value distribution.
\newblock In {\em Entropy-based parameter estimation in hydrology}, pages
  169--183. Springer, 1998.

\bibitem{raynal2021general}
J.~A.~Raynal Villase{\~n}or.
\newblock General extreme value distribution.
\newblock In {\em Frequency Analyses of Natural Extreme Events: A Spreadsheets
  Approach}, pages 233--284. Springer, 2021.

\bibitem{tsiftsi2018extreme}
T.~Tsiftsi and V.~De la~Luz.
\newblock Extreme value analysis of solar flare events.
\newblock {\em Space Weather}, 16(12):1984--1996, 2018.

\bibitem{redner2001guide}
S.~Redner.
\newblock {\em A guide to first-passage processes}.
\newblock Cambridge University Press, 2001.

\bibitem{TargetSearch2024}
D.~Grebenkov, R.~Metzler, and G.~Oshanin.
\newblock {\em Target Search Problems}.
\newblock Springer International Publishing AG, 2024.

\bibitem{benichou2014first}
O.~B{\'e}nichou and R.~Voituriez.
\newblock From first-passage times of random walks in confinement to
  geometry-controlled kinetics.
\newblock {\em Physics Reports}, 539(4):225--284, 2014.

\bibitem{godec2016universal}
A.~Godec and R.~Metzler.
\newblock Universal proximity effect in target search kinetics in the
  few-encounter limit.
\newblock {\em Physical Review X}, 6(4):041037, 2016.

\bibitem{bray2013persistence}
A.~J. Bray, S.~N. Majumdar, and G.~Schehr.
\newblock Persistence and first-passage properties in nonequilibrium systems.
\newblock {\em Advances in Physics}, 62(3):225--361, 2013.

\bibitem{meyer2011universality}
B.~Meyer, C.~Chevalier, R.~Voituriez, and O.~B{\'e}nichou.
\newblock Universality classes of first-passage-time distribution in confined
  media.
\newblock {\em Physical Review E}, 83(5):051116, 2011.

\bibitem{grebenkov2023boundary}
D.~S. Grebenkov and A.~T. Skvortsov.
\newblock Boundary homogenization for target search problems.
\newblock {\em arXiv preprint arXiv:2310.14322}, 2023.

\bibitem{scher2023escape}
Y.~Scher, S.~Reuveni, and D.~S. Grebenkov.
\newblock Escape of a sticky particle.
\newblock {\em Physical Review R}, 5(4):043196, 2023.

\DIFaddbegin \bibitem{singh2025sokoban}
\DIFadd{P.~Singh, D.~A. Kessler, and E.~Barkai.
}\newblock \DIFadd{Sokoban random walk: From environment reshaping to trapping transition.
}\newblock {\em \DIFadd{arXiv preprint arXiv:2508.07825}}\DIFadd{, 2025.
}


\DIFaddend \bibitem{lawley2024competition}
S.~D. Lawley.
\newblock Competition of many searchers.
\newblock In {\em Target Search Problems}, pages 281--303. Springer, 2024.




\bibitem{grebenkov2022first}
D.~S. Grebenkov and A.~Kumar.
\newblock First-passage times of multiple diffusing particles with reversible
  target-binding kinetics.
\newblock {\em Journal of Physics A: Mathematical and Theoretical},
  55(32):325002, 2022.

\bibitem{basnayake2019asymptotic}
K.~Basnayake, Z.~Schuss, and D.~Holcman.
\newblock Asymptotic formulas for extreme statistics of escape times in 1, 2
  and 3-dimensions.
\newblock {\em Journal of Nonlinear Science}, 29:461--499, 2019.

\bibitem{lawley2020distribution}
S.~D. Lawley.
\newblock Distribution of extreme first passage times of diffusion.
\newblock {\em Journal of Mathematical Biology}, 80(7):2301--2325, 2020.

\bibitem{lawley2020extreme}
S.~D. Lawley.
\newblock Extreme first-passage times for random walks on networks.
\newblock {\em Physical Review E}, 102(6):062118, 2020.

\bibitem{linn2022extreme}
S.~Linn and S.~D. Lawley.
\newblock Extreme hitting probabilities for diffusion.
\newblock {\em Journal of Physics A: Mathematical and Theoretical},
  55(34):345002, 2022.

\bibitem{ellettari2025rare}
E.~Ellettari, G.~Nasuti, A.~Bassanoni, A.~Vezzani, and R.~Burioni.
\newblock Rare events, many searchers, and fast target reaching in a finite
  domain.
\newblock {\em arXiv preprint arXiv:2507.09452}, 2025.

\bibitem{meerson2015mortality}
B.~Meerson and S.~Redner.
\newblock Mortality, redundancy, and diversity in stochastic search.
\newblock {\em Physical review letters}, 114(19):198101, 2015.

\DIFaddbegin \bibitem{franke2012survival}
\DIFadd{J.~Franke and S.~N.~Majumdar.
}\newblock \DIFadd{Survival probability of an immobile target surrounded by mobile traps.
}\newblock {\em \DIFadd{Journal of Statistical Mechanics: Theory and Experiment}}\DIFadd{, 2012(05):P05024, 2012.
}


\bibitem{mejia2011first} 
C.~Mej\'ia-Monasterio, G.~Oshanin, and G.~Schehr.
\newblock First passages for a search by a swarm of independent random searchers.
\newblock {\em Journal of Statistical Mechanics: Theory and Experiment},
2011(06):P06022, 2011.



\bibitem{madrid2020competition}
J.~B. Madrid and S.~D. Lawley.
\newblock Competition between slow and fast regimes for extreme first passage
  times of diffusion.
\newblock {\em Journal of Physics A: Mathematical and Theoretical},
  53(33):335002, 2020.

\bibitem{basnayake2019fastest}
K.~Basnayake and D.~Holcman.
\newblock Fastest among equals: a novel paradigm in biology. reply to comments:
  Redundancy principle and the role of extreme statistics in molecular and
  cellular biology.
\newblock {\em Physics of life reviews}, 28:96--99, 2019.

\bibitem{bao2006last}
J.~D. Bao and Y.~Jia.
\newblock Last passage time statistics for barrier-crossing processes.
\newblock {\em Journal of statistical physics}, 123:861--869, 2006.



\bibitem{grebenkov2022reversible}
D.~S. Grebenkov and A.~Kumar.
\newblock Reversible target-binding kinetics of multiple impatient particles.
\newblock {\em The Journal of Chemical Physics}, 156(8), 2022.

\bibitem{ghodsi1991performance}
M.~Ghodsi and K.~Kant.
\newblock Performance analysis of parallel search algorithms on multiprocessor
  systems.
\newblock {\em Performance evaluation}, 13(1):67--81, 1991.

\bibitem{dey2019dissipative}
K.~K. Dey and A.~Sen.
\newblock The design of dissipative molecular assemblies driven by reaction
  cycles.
\newblock {\em Current Opinion in Colloid \& Interface Science}, 39:85--97,
  2019.

\bibitem{schrodinger1915theorie}
E.~Schr{\"o}dinger.
\newblock Zur theorie der fall-und steigversuche an teilchen mit brownscher
  bewegung.
\newblock {\em Physikalische Zeitschrift}, 16:289--295, 1915.

\bibitem{majumdar1999persistence}
S.~N. Majumdar.
\newblock Persistence in nonequilibrium systems.
\newblock {\em Current Science}, pages 370--375, 1999.

\bibitem{havlin2002diffusion}
S.~Havlin and D.~Ben-Avraham.
\newblock Diffusion in disordered media.
\newblock {\em Advances in physics}, 51(1):187--292, 2002.

\bibitem{comtet2020last}
A.~Comtet, F.~Cornu, and G.~Schehr.
\newblock Last-passage time for linear diffusions and application to the
  emptying time of a box.
\newblock {\em Journal of Statistical Physics}, 181(5):1565--1602, 2020.

\bibitem{meinecke2017multiscale}
L.~Meinecke.
\newblock Multiscale modeling of diffusion in a crowded environment.
\newblock {\em Bulletin of mathematical biology}, 79(11):2672--2695, 2017.

\bibitem{baravi2025first}
T.~Baravi, D.~A. Kessler, and E.~Barkai.
\newblock First passage times in compact domains exhibit biscaling.
\newblock {\em Physical Review Letters}, 134(12):127101, 2025.

\bibitem{feller1971introduction}
W.~Feller.
\newblock An introduction to probability theory and its applications.
\newblock 1971.

\bibitem{leadbetter2012extremes}
M.~R. Leadbetter, G.~Lindgren, and H.~Rootz{\'e}n.
\newblock {\em Extremes and related properties of random sequences and
  processes}.
\newblock Springer Science \& Business Media, 2012.

\bibitem{eliazar2019poisson}
I.~Eliazar, R.~Metzler, and S.~Reuveni.
\newblock Poisson-process limit laws yield gumbel max-min and min-max.
\newblock {\em Physical Review E}, 100(2):022129, 2019.

\bibitem{baravi2025solutions}
T.~Baravi, D.~A. Kessler, and E.~Barkai.
\newblock Solutions of first-passage time problems: A biscaling approach.
\newblock {\em Physical Review E}, 111(4):044103, 2025.

\bibitem{ben2000diffusion}
D.~Ben-Avraham and S.~Havlin.
\newblock {\em Diffusion and reactions in fractals and disordered systems}.
\newblock Cambridge university press, 2000.

\bibitem{gefen1983anomalous}
Y.~Gefen, A.~Aharony, and S.~.
\newblock Anomalous diffusion on percolating clusters.
\newblock {\em Physical Review Letters}, 50(1):77, 1983.

\bibitem{rammal1983random}
R.~Rammal and G.~Toulouse.
\newblock Random walks on fractal structures and percolation clusters.
\newblock {\em Journal de Physique Lettres}, 44(1):13--22, 1983.

\bibitem{bouchaud1990anomalous}
J.-P. Bouchaud and A.~Georges.
\newblock Anomalous diffusion in disordered media: statistical mechanisms,
  models and physical applications.
\newblock {\em Physics reports}, 195(4-5):127--293, 1990.




\bibitem{Note1}
In the classification of Metzler and Nonnenmacher, the OP case uses the
  similarity variable $\xi =r\protect \,t^{-1/d_w}$ with tail parameters
  $\alpha =0$ and $u=d_w$ [85].

\bibitem{teguia2005sierpi}
A.~M. Teguia and A.~P. Godbole.
\newblock Sierpiński gasket graphs and some of their properties.
\newblock {\em arXiv preprint math/0509259}, 2005.
\DIFaddbegin 

\bibitem{leibovich2013everlasting}\DIFadd{N.~Leibovich and E.~Barkai.
}\newblock \DIFadd{Everlasting effect of initial conditions on single-file diffusion.
}\newblock {\em \DIFadd{Physical Review E}}\DIFadd{, 88(3):032107, 2013.
}



\bibitem{ben2005transition}
\DIFadd{G.~Ben~Arous, S.~Molchanov, and A.~F. Ramírez.
}\newblock \DIFadd{Transition from the annealed to the quenched asymptotics for a random walk on random obstacles.
}\newblock {\em \DIFadd{Annals of Probability}}\DIFadd{, 33(6):2149--2187, 2005.
}

\bibitem{Donsker1975}
\DIFadd{M.~D. Donsker and S.~R.~S. Varadhan.
}\newblock \DIFadd{Asymptotics for the Wiener sausage.
}\newblock {\em \DIFadd{Communications on Pure and Applied Mathematics}}\DIFadd{, 28(4):525--565, 1975.
}

\bibitem{Donsker1979}
\DIFadd{M.~D. Donsker and S.~R.~S. Varadhan.
}\newblock \DIFadd{On the number of distinct sites visited by a random walk.
}\newblock {\em \DIFadd{Communications on Pure and Applied Mathematics}}\DIFadd{, 32(6):721--747, 1979.
}



\bibitem{Hartmann2023}
\DIFadd{A.~K. Hartmann, S.~N. Majumdar, and G.~Schehr.
}\newblock \DIFadd{Distribution of the maximum of independent resetting Brownian motions.
}\newblock {\em \DIFadd{Journal of Statistical Mechanics: Theory and Experiment}}\DIFadd{, 2023(9):093205, 2023.
}

\bibitem{Burenev2024}
\DIFadd{I.~N. Burenev, M.~V. Tamm, and S.~N. Majumdar.
}\newblock \DIFadd{Occupation time of a system of Brownian particles on the line: large deviations for annealed and quenched averages.
}\newblock {\em \DIFadd{Physical Review E}}\DIFadd{, 109(4):044150, 2024.
}

\bibitem{Madrid2020}
\DIFadd{J.~B. Madrid and S.~D. Lawley.
}\newblock \DIFadd{Competition between slow and fast regimes for extreme first passage times of diffusion.
}\newblock {\em \DIFadd{Journal of Physics A: Mathematical and Theoretical}}\DIFadd{, 53(33):335002, 2020.
}\DIFaddend 



\bibitem{o1985analytical}
B.~O'Shaughnessy and I.~Procaccia.
\newblock Analytical solutions for diffusion on fractal objects.
\newblock {\em Physical Review Letters}, 54(5):455, 1985.

\bibitem{o1985diffusion}
B.~O’Shaughnessy and I.~Procaccia.
\newblock Diffusion on fractals.
\newblock {\em Physical Review A}, 32(5):3073, 1985.

\bibitem{Note2}
OP can be viewed as the $\nu =1$ limit of the unified nonlinear
  anomalous-diffusion equation of Pedron \protect \textit {et al.}~[86].

\bibitem{metzler1997fractional}
R.~Metzler and T.~F. Nonnenmacher.
\newblock Fractional diffusion: exact representations of spectral functions.
\newblock {\em Journal of Physics A: Mathematical and General}, 30(4):1089,
  1997.

\bibitem{pedron2002nonlinear}
I.~T. Pedron, R.~S. Mendes, L.~C. Malacarne, and E.~K. Lenzi.
\newblock Nonlinear anomalous diffusion equation and fractal dimension: Exact
  generalized gaussian solution.
\newblock {\em Physical Review E}, 65(4):041108, 2002.

\end{thebibliography}

\DIFaddend

\end{document}